\documentclass[journal,twocolumn,10pt,twoside]{IEEEtranTCOM}

\normalsize

\ifCLASSINFOpdf
\else
\fi

\usepackage{amsmath}
\usepackage{amssymb}
\usepackage{cite}
\usepackage{graphicx}
\usepackage{subfigure}
\usepackage{multirow} 
\usepackage{longtable}
\usepackage{threeparttable}
\usepackage{caption3}
\usepackage{array}
\usepackage{cases}
\usepackage{bbm}
\usepackage{color}

\newtheorem{theorem}{Theorem}

\newcommand\mycom[2]{\genfrac{}{}{0pt}{}{#1}{#2}}

\begin{document}
\title{Space-Time Encoded MISO Broadcast Channel with Outdated CSIT: An Error Rate and Diversity Performance Analysis}

\author{Bruno Clerckx and David Gesbert
\thanks{Bruno Clerckx is with Imperial College London, London SW7 2AZ, United Kingdom and also with the School
of Electrical Engineering, Korea University, Seoul 136-701, Korea~(email: b.clerckx@imperial.ac.uk). David Gesbert is with EURECOM,
Sophia-Antipolis, France (email: david.gesbert@eurecom.fr). This work was supported in part by the Seventh Framework Programme for Research of the European Commission under grant number HARP-318489.}}

\markboth{IEEE Transactions on Communications}%
{Submitted paper}

\maketitle

\begin{abstract}
Studies of the MISO Broadcast Channel (BC) with delayed
Channel State Information at the Transmitter (CSIT) have so far focused on the
sum-rate and Degrees-of-Freedom (DoF) region analysis. In this
paper, we investigate for the first time the error rate performance
at finite SNR and the diversity-multiplexing tradeoff
(DMT) at infinite SNR of a space-time encoded transmission
over a two-user MISO BC with delayed CSIT. We
consider the so-called MAT protocol obtained by Maddah-Ali and
Tse, which was shown to provide 33\% DoF enhancement over
TDMA. While the asymptotic DMT analysis shows that MAT is
always preferable to TDMA, the Pairwise Error Probability analysis at finite SNR shows
that MAT is in fact not always a better alternative to TDMA.
Benefits can be obtained over TDMA only at very high rate or
once concatenated with a full-rate full-diversity space-time code.
The analysis is also extended to spatially correlated
channels and the influence of transmit correlation matrices and
user pairing strategies on the performance are discussed. Relying
on statistical CSIT, signal constellations are further optimized
to improve the error rate performance of MAT and make it
insensitive to user orthogonality. Finally, other transmission strategies relying on delayed
CSIT are discussed.
\end{abstract}


\IEEEpeerreviewmaketitle

\section{Introduction}

\IEEEPARstart{T}{he} performance of multiuser MISO/MIMO highly depends on the accuracy of Channel State Information at the Transmitter (CSIT) \cite{Clerckx:2013,Gesbert:2007}. Obtaining accurate CSIT is however challenging in practice as the feedback is subject to various impairments including delay. There has been recent progress on understanding how perfect but outdated CSIT can be beneficial to boost the Degrees-of-Freedom (DoF) of MISO Broadcast Channels \cite{MAT:2012}. It was shown that in a two-user MISO BC with outdated CSIT, a sum DoF of 4/3 can be achieved using a transmission strategy that exploits the knowledge of the outdated CSIT to reconstruct and transmit the interference overheard by co-scheduled receivers. This leads to a 33\% DoF enhancement compared to conventional TDMA approach. Throughout the paper, we will denote such transmission strategy as MAT. In \cite{Kobayashi:2012}, then in \cite{Yang:2013,Gou:2012}, authors extended the work by looking at a scenario with both imperfect instantaneous CSIT and perfect delayed CSIT where the idea of imperfect current CSIT lies in the fact that prediction can be applied to delayed CSIT. An alternative transmission strategy, denoted as Alternative MAT in this paper, was suggested to also achieve a sum DoF of 4/3 and used as a building block of a more general strategy suitable for time correlated channels. 
\cite{Tandon:2013} generalized the findings in \cite{MAT:2012} by giving an optimal DoF region for an alternative CSIT setting in a two-user, where the CSIT of each user can be perfect, delayed or absent. In \cite{Chen:2012a}, the results of \cite{Yang:2013,Gou:2012} were extended by considering different qualities of instantaneous CSIT of the two users. Further analysis on the impact of imperfect outdated CSIT on the DoF region was made in \cite{Chen:2012b}. Recently, DoF results found in the two-user time-correlated MISO BC with delayed CSIT have been extended to the MIMO and interference channel cases in \cite{Yi:2013b}. 
So far, all cited works exclusively focused on DoF analysis at high SNR. In \cite{Yi:2013a}, a sum-rate criterion is considered and the design of precoders to enhance the performance of MAT at finite SNR is addressed. In \cite{Wang:2013,Mingbo:2014}, a similar criterion is chosen but the precoders are designed based on statistical CSIT. In \cite{Chen:2012}, authors depart from classical DoF and rate analysis and also investigate diversity performance of MAT strategy. An outer-bound on the asymptotic diversity-multiplexing tradeoff (DMT) at infinite SNR achieved by MAT is derived and a novel scheme is proposed to achieve both full DoF as well as full diversity.
\par Surprisingly, several transmission strategies have been derived to cope with delayed CSIT in the cited references and much is known about their DoF performance but the diversity performance (including the asymptotic DMT) has been overlooked so far. In particular, the error rate performance of corresponding techniques at finite SNR in MISO BC with outdated CSIT has never been addressed so far. The objective of this paper is to fill this gap by contrasting the asymptotic DMT behavior and the finite SNR error rate performance and identify whether gains over the scenario where delayed CSIT is simply ignored (e.g.\ TDMA) are always observed. Specifically, the contributions of the paper are listed as follows:
\begin{itemize}
\item For a two-user MISO BC, the average Pairwise Error Probability (PEP) over spatially correlated (at the transmitter) Rayleigh fading channels is derived for a general space-time encoded MAT strategy. The PEP behaviour is discussed as a function of the SNR and the transmit correlation matrices of both users and contrasted with conventional behaviours of multi-user MIMO with perfect CSIT and point-to-point MIMO. 
\item The exact diversity-multiplexing tradeoff (DMT) and the one achieved by various space-time encoded MAT in i.i.d.\ Rayleigh fading channels are identified and a suitable space-time code design criterion is proposed. Interestingly, the exact DMT of MAT is shown to match the upper bound found in \cite{Chen:2012}.
\item Relying on the PEP analysis, a suitable user pairing strategy is discussed. Pairing statistically orthogonal users with similar magnitudes of the transmit correlation coefficients is shown to be a good strategy. With such a strategy, transmit correlation is shown to have a minor impact on the error rate performance of MAT over a large range of SNR.

\item If transmit spatial correlations are known at the transmitter, following a previous work by the authors \cite{Clerckx:2004}, signal constellations are further optimized to improve the error rate performance of MAT in spatially correlated channels. The performance with such constellations are shown to be insensitive to user orthogonality.
\item The error rate and DMT performance analysis is finally extended to other strategies relying on delayed CSIT, namely the Alternative MAT proposed in \cite{Yang:2013}. Despite the fact that both MAT and Alternative MAT have the same mutliplexing gain (DoF), Alternative MAT is shown to incur lower diversity and coding gains than the original MAT. 
\end{itemize}

\par Overall, the primary takeaway observation from this work is that in i.i.d. Rayleigh fading channels, MAT always outperforms TDMA (that would ignore the delayed CSIT) from an asymptotic DMT perspective but is not always a better alternative to TDMA from an error rate perspective at finite SNR. Benefits are obtained over TDMA only at very high rate (no gains have been identified at a per-user rate of 3 bit/s/Hz or below) or once concatenated with a full-rate full-diversity space-time code. Alternative MAT, while being superior to TDMA from an asymptotic DMT perspective, is on the other hand outperformed by TDMA over a wide SNR and rate range when it comes to error probability performance. In spatially correlated channels, MAT with a suitable user pairing strategy is shown to be less sensitive to transmit spatial correlation than TDMA.

\par The paper is organized as follows. Section \ref{system_model} details the system model. Section \ref{iid_channel} details the error rate performance in i.i.d.\ Rayleigh fading channels and discusses the diversity-multiplexing tradeoff and code design criterion. Section \ref{correlated_channel} extends the analysis to spatially correlated channels and derives user pairing strategies and optimized signal constellations. Section \ref{Section_Alt_MAT} extends the discussions to alternative schemes and Section \ref{evaluations} confirms through simulations the observations made from the analytical derivations. Section \ref{conclusions} concludes the work.

\par The following notations are used throughout the paper. Bold lower case and upper case letters stand for vectors and matrices respectively whereas a symbol not in bold font represents a scalar. $\left(.\right)^T$ and $\left(.\right)^H$ represent the transpose and conjugate transpose of a matrix or vector respectively. Operators $\det\left(.\right)$ and $\textnormal{Tr}\left\{.\right\}$ refer to the determinant and trace of a matrix, respectively. $\mathcal{E\left\{.\right\}}$ refers to the expectation operator. Logarithm $\log$ is taken in base $e$, unless otherwise specified. $\textnormal{diag}\left\{\mathbf{a}\right\}$ refers to the diagonal matrix whose elements are given by the entries of $\mathbf{a}$. Finally, $f(\rho)\stackrel{.}{=}g(\rho)$ indicates $\lim_{\rho\rightarrow \infty}\frac{f(\rho)}{g(\rho)}=1$. $\stackrel{.}{\leq}$ is defined similarly.

\section{System Model}\label{system_model}
Let us assume a two-user two transmit antenna MISO BC with delayed CSIT. The transmission occurs over three coherence times. Each coherence time is made of $T$ time slots over which the channel is constant. We denote the channel vector of user 1 on coherence time $k$ as $\mathbf{h}_k=\left[\begin{array}{cc}h_{k,1} & h_{k,2}\end{array}\right]$ where any entry $h_{k,m}$ refers to the channel coefficient from transmit antenna $m$. Similarly, $\mathbf{g}_k=\left[\begin{array}{cc}g_{k,1} & g_{k,2}\end{array}\right]$ and $g_{k,m}$ are defined for user 2. 

\par Denoting the transmit signal on time slot $t$ of coherence time $k$ as $\mathbf{x}_{k,t}$, the received signals at user 1 and 2, respectively denoted as $y_{k,t}$ and $z_{k,t}$, write as
\begin{align}
y_{k,t}&=\mathbf{h}_k \mathbf{x}_{k,t}+n_{k,t},\\
z_{k,t}&=\mathbf{g}_k \mathbf{x}_{k,t}+w_{k,t},
\end{align}
where $n_{k,t} \sim \mathcal{CN}(0,1)$ and $w_{k,t}\sim \mathcal{CN}(0,1)$ are AWGN. We consider a long-term power constraint $\mathcal{E}\big\{\mathbf{x}_{k,t}^H\mathbf{x}_{k,t}\big\}\leq \rho_{lt}$.

\par The channel coefficients are modeled as identically distributed circularly symmetric complex Gaussian variables but can be either independent as in Section \ref{iid_channel} or spatially correlated as in Section \ref{correlated_channel}. They are assumed constant within a coherence time and change independently from one coherence time to the next one. The CSI is assumed to be available at the transmitter only at the next coherence time. For simplicity, the normalization $\mathcal{E}\big\{\left|h_{k,m}\right|^2\big\}=\mathcal{E}\big\{\left|g_{k,m}\right|^2\big\}=1$ is made. We define the transmit covariance matrices $\mathbf{R}_{t,1}$ and $\mathbf{R}_{t,2}$ for user 1 and 2 respectively as
\begin{align}\label{correlation_matrices}
\begin{split}
\mathbf{R}_{t,1}&=\mathcal{E}\left\{\mathbf{h}_k^{H}\mathbf{h}_k\right\}=\left[\begin{array}{cc}1 & t_1^{*} \\ t_1 & 1 \end{array}\right], \hspace{0.5cm} \forall k\\ 
\mathbf{R}_{t,2}&=\mathcal{E}\left\{\mathbf{g}_k^{H}\mathbf{g}_k\right\}=\left[\begin{array}{cc}1 & t_2^{*} \\ t_2 & 1 \end{array}\right], \hspace{0.5cm} \forall k.
\end{split}
\end{align}
The quantities $t_1$ and $t_2$ are the transmit correlation coefficients and can be expressed in terms of their magnitudes and phases as $t_1=\left|t_1\right|e^{j\varphi_1}$ and $t_2=\left|t_2\right|e^{j\varphi_2}$.
Whenever we assume i.i.d.\ circularly symmetric complex Gaussian variables (denoted in short as i.i.d.\ Rayleigh fading in the sequel), $\mathbf{R}_{t,1}$ and $\mathbf{R}_{t,2}$ are identity matrices.

\subsection{From MAT to Space-Time Encoded MAT}
\par  We consider two independent codewords, $\mathbf{C}=\left[\mathbf{c}_{1},\ldots,\mathbf{c}_{T}\right]$ and $\mathbf{C}'=\left[\mathbf{c}'_{1},\ldots,\mathbf{c}'_{T}\right]$ respectively intended for user 1 and user 2. Their sizes are $2 \times T$, therefore spanning the two transmit antennas and $T$ time slots. The codewords are normalized such that $\mathcal{E}\left\{\textnormal{Tr}\left\{\mathbf{C}\mathbf{C}^H\right\}\right\}=\mathcal{E}\left\{\textnormal{Tr}\left\{\mathbf{C}'\mathbf{C}'^H\right\}\right\}=T$. The aim is to transmit codeword $\mathbf{C}$ to user $1$ and codeword $\mathbf{C}'$ to user $2$ over $3T$ time slots using the MAT strategy\cite{MAT:2012}. The system can therefore be viewed as a space-time encoded transmission over a two-user MISO BC with delayed CSIT. For simplicity, we will look at the performance of the first user only. 
\par In MAT, it is commonly assumed that the codewords $\mathbf{C}$ and $\mathbf{C}'$ are chosen as in Spatial Multiplexing (SM), i.e.\ independent data streams are transmitted from each transmit antenna, because the focus is on DoF maximization. Assuming an uncoded (no FEC) transmission, the time slot index $t$ can be dropped and the $2\times 1$ codewords $\mathbf{C}=\mathbf{c}$ and $\mathbf{C}'=\mathbf{c}'$ span only one symbol duration ($T=1$). As detailed in \cite{MAT:2012}, MAT transmits $\mathbf{x}_{1}=\sqrt{\rho}\mathbf{c}$ in coherence time 1, $\mathbf{x}_{2}=\sqrt{\rho}\mathbf{c}'$ in coherence time 2 and the overheard interference $\mathbf{x}_{3}=\sqrt{\rho}\big[\begin{array}{cc}\mathbf{g}_1 \mathbf{c}+\mathbf{h}_2\mathbf{c}' & 0 \end{array}\big]^T$ in coherence time 3. A long-term average transmit power (where averaging is also taken over the channel realizations) of $\rho_{lt}=4/3\rho$ is consumed and twice as much power is spent on coherence time 3 as in coherence time 1 and 2. In the sequel, we will refer to $\rho$ as the SNR. It results in the following equivalent system model for user 1 
\begin{equation}
\left[\begin{array}{c}y_{1}\\y_{2}\\y_{3}\end{array}\right]=\sqrt{\rho}\left[\begin{array}{c} \mathbf{h}_{1} \\ \mathbf{0} \\ h_{3,1}\mathbf{g}_{1}\end{array}\right]\mathbf{c}+\sqrt{\rho}\left[\begin{array}{c} \mathbf{0} \\ \mathbf{h}_{2} \\ h_{3,1}\mathbf{h}_{2}\end{array}\right]\mathbf{c}'+\left[\begin{array}{c}n_{1}\\n_{2}\\n_{3}\end{array}\right]. 
\end{equation}
After further interference elimination,
\begin{equation}
\tilde{\mathbf{y}}=\left[\begin{array}{c}y_{1}\\y_{3}-h_{3,1} y_{2}\end{array}\right]=\sqrt{\rho}\mathbf{H}\mathbf{c}+\left[\begin{array}{c}n_{1}\\n_{3}-h_{3,1} n_{2}\end{array}\right]  
\end{equation}
where
\begin{equation}\label{H_system_model_MAT}
\mathbf{H}=\left[\begin{array}{cc} h_{1,1} & h_{1,2} \\ h_{3,1}g_{1,1} & h_{3,1} g_{1,2}\end{array}\right].
\end{equation}
This is an equivalent $2\times2$ MIMO channel and with the instantaneous channel realizations perfectly known at the receive side and shared across users (as $\mathbf{H}$ is a function of both users' channels), an estimate of the entries of $\mathbf{c}$ can be obtained using e.g.\ a Maximum-Likelihood (ML) or linear receiver. This transmission strategy is denoted as SM-encoded MAT in the sequel.

\par However, MAT as presented in \cite{MAT:2012} is a framework that is applicable to any number $T$ of time slots and is not limited to SM-type of codewords. 
Hence, in the rest of this paper, we will often refer to O-STBC encoded MAT or more generally space-time encoded MAT to stress that the codewords $\mathbf{C}$ and $\mathbf{C}'$ are either chosen as O-STBC or from a general space-time code (including SM-encoded MAT). On time slot $t$, a space-time encoded MAT consists in transmitting $\mathbf{x}_{1,t}=\sqrt{\rho}\mathbf{c}_{t}$ in coherence time 1, $\mathbf{x}_{2,t}=\sqrt{\rho}\mathbf{c}'_{t}$ in coherence time 2 and the overheard interference $\mathbf{x}_{3,t}=\sqrt{\rho}\big[\begin{array}{cc}\mathbf{g}_1 \mathbf{c}_{t}+\mathbf{h}_2\mathbf{c}'_{t} & 0 \end{array}\big]^T$ in coherence time 3. The equivalent system model for user 1 at time instant $t=1,\ldots,T$ can then be written as 
\begin{equation}
\left[\begin{array}{c}y_{1,t}\\y_{2,t}\\y_{3,t}\end{array}\right]=\sqrt{\rho}\left[\begin{array}{c} \mathbf{h}_{1} \\ \mathbf{0} \\ h_{3,1} \mathbf{g}_{1}\end{array}\right]\mathbf{c}_{t}+\sqrt{\rho}\left[\begin{array}{c} \mathbf{0} \\ \mathbf{h}_{2} \\ h_{3,1} \mathbf{h}_{2}\end{array}\right]\mathbf{c}'_{t}+\left[\begin{array}{c}n_{1,t}\\n_{2,t}\\n_{3,t}\end{array}\right], 
\end{equation}
and after further interference elimination
\begin{equation}\label{system_model_MAT}
\tilde{\mathbf{y}}_{t}=\left[\begin{array}{c}y_{1,t}\\y_{3,t}-h_{3,1} y_{2,t}\end{array}\right]=\sqrt{\rho}\mathbf{H}\mathbf{c}_{t}+\left[\begin{array}{c}n_{1,t}\\n_{3,t}-h_{3,1} n_{2,t}\end{array}\right].  
\end{equation}
\par Users perceive a space-time encoded transmission over an equivalent $2\times2$ MIMO channel. We note the difference in terms of channel matrix $\mathbf{H}$ in \eqref{H_system_model_MAT} and noise compared to a classical $2\times2$ point-to-point space-time encoded MIMO system model \cite{Tarokh:1998}. In particular, the entries of the channel matrix $\mathbf{H}$ are not identically distributed and are function of both users' channels. Those differences make the performance analysis (including the error rate, diversity-multiplexing tradeoff, impact of spatial correlation) different from conventional point-to-point MIMO Rayleigh fading channels and multi-user MISO with perfect CSIT.

\subsection{Performance Metrics}

\par Since the encoded transmission is performed over a small number of channel realizations and the transmitter does not have perfect CSIT, outage may occur. Hence outage and error probabilities are valid performance metrics that will be used throughout this paper. The outage probability is used to identify the asymptotic DMT and the error probability is used to characterize the behavior of practical space-time encoded MAT strategies at finite SNR.
\par It is assumed that ML decoding is performed at receiver 1 (and similarly for receiver 2) in order to estimate $\mathbf{C}$ from the received signals $\tilde{\mathbf{y}}_{t}$, $t=1,\ldots,T$. With instantaneous channel realizations perfectly known at the receive side and shared across users, the ML decoder at receiver 1 computes an estimate of the transmitted codeword according to
\begin{equation}
	\hat{\mathbf{C}}=\text{arg} \min_{\mathbf{C}}\sum_{t=1}^{T}\left\|\mathbf{\Sigma}^{-1/2}\left(\tilde{\mathbf{y}}_t-\sqrt{\rho}\mathbf{H}\mathbf{c}_t\right)\right\|^2
	\label{equation2}
\end{equation}
where $\mathbf{\Sigma}=\textnormal{diag}\big\{1,1+\left|h_{3,1}\right|^2\big\}$ is the covariance matrix of the noise vector in \eqref{system_model_MAT}. The minimization in \eqref{equation2} is performed over all possible codewords. 

\par Let us define the following matrix $\tilde{\mathbf{H}}$ 
\begin{equation}
\tilde{\mathbf{H}}=\mathbf{\Sigma}^{-1/2}\mathbf{H}=\left[\begin{array}{cc}1 & 0 \\ 0 & X\end{array}\right]\underbrace{\left[\begin{array}{cc} h_{1,1} & h_{1,2} \\ g_{1,1} & g_{1,2}\end{array}\right]}_{\mathbf{H}'}
\end{equation}
where $X=\frac{h_{3,1}}{\sqrt{1+\left|h_{3,1}\right|^2}}$. When a codeword $\mathbf{C}$ is transmitted, we are interested in the error probability, called Pairwise Error Probability (PEP), that the ML decoder decodes the codeword $\mathbf{E}$ instead of $\mathbf{C}$. The conditional PEP $P\big(\mathbf{C}\rightarrow \mathbf{E}\left|\right.\tilde{\mathbf{H}}\big)$ with a ML decoder can then be written as 
\begin{equation}
P\left(\mathbf{C}\rightarrow \mathbf{E}\left|\right.\tilde{\mathbf{H}}\right)=\mathcal{Q}\left(\sqrt{\frac{\rho}{2}\left\|\tilde{\mathbf{H}}\left(\mathbf{C}-\mathbf{E}\right)\right\|_F^2}\right).
\end{equation}
Even though the analysis could be conducted based on the exact PEP (by making use of the Craig's formula \cite{Clerckx:2013}), for the sake of readability and simplicity, we will make use of the Chernoff bound and therefore upper bound the PEP as
\begin{equation}\label{MAT_PEP_cond}
P\left(\mathbf{C}\rightarrow \mathbf{E}\left|\right.\tilde{\mathbf{H}}\right)\leq \exp\left(-\frac{\rho}{4}\left\|\tilde{\mathbf{H}}\left(\mathbf{C}-\mathbf{E}\right)\right\|_F^2\right).
\end{equation}

\section{Uncorrelated Fading Channels}\label{iid_channel}
We discuss spatially uncorrelated and correlated scenarios in two different sections. We start here with the uncorrelated case, i.e.\ $\mathbf{R}_{t,1}=\mathbf{R}_{t,2}=\mathbf{I}_2$.
The average PEP for both MAT is obtained by taking the expectation of the conditional PEP \eqref{MAT_PEP_cond} over the channel distribution. The expectation can be computed in two steps: first by taking the expectation over the distribution of $\mathbf{H}'$ and then over the distribution of $X$.

\subsection{Error Rate Performance of (encoded) MAT}\label{iid_error_rate_section}
Following the derivations in Appendix \ref{Appendix_MAT}, we get the upper bound on the average PEP of space-time encoded MAT, as displayed in \eqref{MAT_PEP}, by taking the expectation over the channel distribution.
\begin{table*}
\begin{equation}\label{MAT_PEP}
P\left(\mathbf{C}\rightarrow \mathbf{E}\right)\leq\frac{1}{b_1 b_2}\left[\frac{1}{b_1 b_2}+\frac{\left(b_2-1\right)^2}{b_2^2\left(b_1-b_2\right)}\exp\left(\frac{1}{b_2}\right)\textnormal{Ei}\left(\frac{-1}{b_2}\right)-\frac{\left(b_1-1\right)^2}{b_1^2\left(b_1-b_2\right)}\exp\left(\frac{1}{b_1}\right)\textnormal{Ei}\left(\frac{-1}{b_1}\right)\right]
\end{equation}
\hrulefill
\end{table*}
$\textnormal{Ei}\left(x\right)$ is the exponential integral and $b_k=1+a_k$ ($k=1,2$) with $a_k=\frac{\rho}{4}\lambda_k$ and $\lambda_k$ the $k^{th}$ eigenvalue of the error matrix $\tilde{\mathbf{E}}=\left(\mathbf{C}-\mathbf{E}\right)\left(\mathbf{C}-\mathbf{E}\right)^H$.

\par Assuming a full-rank code, i.e.\ $\lambda_k>0$ for $k=1,2$, the average PEP at sufficiently high SNR can be approximated as
\begin{equation}
P\left(\mathbf{C}\rightarrow \mathbf{E}\right)\leq\left(\frac{\rho}{4}\right)^{-3}\left(\lambda_1 \lambda_2\right)^{-1}\frac{\log\left(\lambda_1\right)-\log\left(\lambda_2\right)}{\lambda_1-\lambda_2}.\label{MAT_PEP_highSNR}
\end{equation}
The maximum achievable diversity gain is 3 and the coding gain is given by $\left(\lambda_1 \lambda_2\right)^{-1}\frac{\log\left(\lambda_1\right)-\log\left(\lambda_2\right)}{\lambda_1-\lambda_2}$ where $\lambda_1 \lambda_2=\det\big(\tilde{\mathbf{E}}\big)$. We note the difference with the classical rank-determinant criterion in space-time code design over i.i.d.\ Rayleigh fading channels \cite{Tarokh:1998}. Intuitively, a maximum diversity gain of 3 (rather than 4) is achieved with a full-rank code because of the presence of $h_{3,1}$ in both entries of the second row of $\mathbf{H}$ in \eqref{H_system_model_MAT}. This implies that the entries of the second row of \eqref{H_system_model_MAT} do not fade independently abd that an error is likely to occur whenever the three channel coefficients $h_{3,1}$, $h_{1,1}$ and $h_{1,2}$ are in deep fade.

\par We particularize the result to the following two cases of Spatial Multiplexing and Orthogonal-Space Time Block Codes (O-STBC).
\par Let us first assume a SM-encoded MAT where the space-time transmission is operated using Spatial Multiplexing (SM) with independent streams transmitted from each antenna without coding across antennas. Such transmission leads to rank-1 error matrix $\tilde{\mathbf{E}}$ with a unique non-zero eigenvalue $\lambda$. Hence $\lambda_2=0$, $b_2=1$, $b_1=b=1+a$ and $b_1-b_2=a$ with $a=\frac{\rho}{4}\lambda$. Denoting $\mathbf{C}-\mathbf{E}=\frac{1}{\sqrt{2}}\big[\begin{array}{cc} c_0-e_0 & c_1-e_1\end{array}\big]^T$, we get $\lambda=\frac{1}{2}\left[\left|c_0-e_0\right|^2+\left|c_1-e_1\right|^2\right]$. The average PEP \eqref{MAT_PEP} simplifies into
\begin{equation}\label{MAT_PEP_SM}
P\left(\mathbf{C}\rightarrow \mathbf{E}\right)\leq \frac{1}{b}\left[\frac{1}{b}-\frac{a}{b^2}\exp\left(\frac{1}{b}\right)\textnormal{Ei}\left(\frac{-1}{b}\right)\right].
\end{equation}
At high SNR,
\begin{equation}\label{SM_highSNR_PEP_MAT}
P\left(\mathbf{C}\rightarrow \mathbf{E}\right)\leq \frac{\log\left(a\right)}{a^2}=\left(\frac{\rho}{4}\right)^{-2}\lambda^{-2}\log\left(\frac{\rho}{4}\lambda\right).
\end{equation}
Focusing on the worst case PEP (i.e.\ maximum PEP among all possible pairs of codewords $\mathbf{C}$ and $\mathbf{E}$ with $\mathbf{C}\neq\mathbf{E}$), the average error probability at high SNR of SM-encoded MAT can be approximated as
\begin{equation}
P_{SM-MAT}\approx \left(\frac{\rho}{4}\right)^{-2}\lambda_{min}^{-2}\log\left(\frac{\rho}{4}\lambda_{min}\right)
\end{equation}
with $\lambda_{min}=\min_{\mathbf{C}\neq\mathbf{E}}\frac{1}{2}\left[\left|c_0-e_0\right|^2+\left|c_1-e_1\right|^2\right]=\frac{1}{2}d_{min,M}^2$ where $d_{min,M}^2$ refers to the squared minimum distance of a constellation with $2^{\frac{3}{2}R}$ points aiming to achieve a per-user rate $R$ (with two symbols transmitted over three channel uses).
Hence,
\begin{equation}\label{SMMAT_Prob}
P_{SM-MAT}\approx\left(\frac{\rho d_{min,M}^2}{8}\right)^{-2} \log\left(\frac{\rho d_{min,M}^2}{8}\right).
\end{equation}
\par From \eqref{SM_highSNR_PEP_MAT}, the diversity gain writes as $d=-\frac{\partial \log\left(P\left(\mathbf{C}\rightarrow \mathbf{E}\right)\right)}{\partial\log\left(a\right)}=2-\frac{1}{\log\left(a\right)}\approx 2$. Hence in the limit of infinite SNR, a diversity gain of 2 is achievable. Strictly speaking, due to the double Rayleigh distribution of some channel coefficients, the slope of the error probability is not as steep as with the classical Rayleigh distributed MIMO/MISO channel, hence leading to a diversity gain that appears slightly lower than 2 at finite SNR. Intuitively, the diversity gain of 2 comes from the fact that the receiver still has access to two independent observations of the transmitted codeword despite the presence of $h_{3,1}$ in both entries of the second row of \eqref{H_system_model_MAT}. 

\par With O-STBC (Alamouti code) \cite{Tarokh:1999} encoded MAT, $\tilde{\mathbf{E}}=\alpha \mathbf{I}_2$, $b_1=b_2=b=1+a=1+\frac{\rho}{4}\alpha$ and the average PEP is written as
\begin{multline}\label{MAT_PEP_OSTBC}
P\left(\mathbf{C}\rightarrow \mathbf{E}\right)\leq \frac{1}{b^2}\left[\frac{b^3-b^2+b}{b^4}\right.\\
\left.+\frac{\left(1-b\right)^2}{b^4}\exp\left(\frac{1}{b}\right)\textnormal{Ei}\left(\frac{-1}{b}\right)\right],
\end{multline}
which leads at high SNR to
\begin{equation}\label{MAT_PEP_OSTBC_highSNR}
P\left(\mathbf{C}\rightarrow \mathbf{E}\right)\leq \frac{1}{a^2}\left[\frac{1}{a}+\frac{\log\left(a\right)}{a^2}\right]\approx \frac{1}{a^3} = \left(\frac{\rho}{4}\right)^{-3} \alpha^{-3}.
\end{equation}
A diversity gain of 3 is achieved. 

\par We note that a simple TDMA transmission that ignores the delayed CSIT would achieve a diversity gain of 2 by simply transmitting to each user at a time using O-STBC within a coherence time. In an i.i.d.\ Rayleigh fading MISO channel with two transmit antennas, focusing on the worst-case PEP, TDMA with O-STBC transmission for each user at a per-user rate $R$ would lead to an error rate at high SNR
\begin{equation}\label{TDMA_Prob}
P_{TDMA}\approx\left(\frac{\rho_{lt} d_{min,T}^{2}}{8}\right)^{-2}=\left(\frac{4}{3}\frac{\rho d_{min,T}^{2}}{8}\right)^{-2}
\end{equation}
where $d_{min,T}^2$ refers to the squared minimum distance of a constellation with $2^{2R}$ points, aiming to achieve a per-user rate $R$ (with two symbols transmitted per O-STBC block every four channel uses). In \eqref{TDMA_Prob}, $\rho_{lt}$ is used so as to have a fair comparison with MAT under the same consumed average power constraint.
We note the difference and similarities between \eqref{TDMA_Prob} and \eqref{SMMAT_Prob}. To achieve the same per-user rate $R$, TDMA requires a larger constellation size than SM-encoded MAT and its performance is therefore affected by a smaller minimum distance, but SM-encoded MAT error rate slope vs.\ SNR is not as steep as that of TDMA due to the presence of an additional term in \eqref{SMMAT_Prob} that scales with $\log(\rho)$. 
Let us operate TDMA at the SNR $\rho_{T}$ and SM-encoded MAT at the SNR $\rho_{M}$. In order to guarantee $P_{TDMA}=P_{SM-MAT}$, $\rho_{T}$ and $\rho_{M}$ need to satisfy the relationship
\begin{equation}
\left(\frac{4}{3}\frac{\rho_{T} d_{min,T}^{2}}{8}\right)^{-2}=\left(\frac{\rho_{M} d_{min,M}^2}{8}\right)^{-2} \log\left(\frac{\rho_{M} d_{min,M}^2}{8}\right),
\end{equation}
which leads to the following SNR gap $\Delta \rho_{dB}$
\begin{align}\label{delta_rho}
&\Delta \rho_{dB}\nonumber\\
&=10\log_{10}\left(\rho_{M}\right)-10\log_{10}\left(\rho_{T}\right)\nonumber\\
&=10\log_{10}\left(\frac{4}{3}\frac{d_{min,T}^2}{d_{min,M}^2}\right)+\frac{10}{2}\log_{10}\log\left(\frac{\rho_{M} d_{min,M}^2}{8}\right)\nonumber\\
&\approx1.25+10\log_{10}\left(\frac{d_{min,T}^2}{d_{min,M}^2}\right)+\frac{10}{2}\log_{10}\log\left(\frac{\rho_{M} d_{min,M}^2}{8}\right).
\end{align}
The SNR gap $\Delta \rho_{dB}$ is a simple function of the two mechanisms highlighted above: the second term in \eqref{delta_rho} is always negative and relates to the ratio of the minimum distances while the first and third terms are positive and the latter increases with the SNR. At low SNR, $\Delta \rho_{dB}$ may be negative and indicates that SM-encoded MAT may outperform TDMA (i.e.\ $\rho_{T}$ has to be larger than $\rho_{M}$ in order to achieve the same error rate) while as the SNR increases, $\Delta \rho_{dB}$ increases and at some point becomes positive, indicating that TDMA outperforms SM-encoded MAT. Therefore, for a fixed rate transmission, the SNR range where SM-encoded MAT is expected to outperform TDMA is concentrated at low SNR and increases as the rate increases. With a full-rate full-diversity code, MAT exhibits a larger diversity gain than TDMA and therefore always outperforms TDMA at high SNR. We recall that $1.25$ dB gap originates from the fact that to operate at an SNR $\rho$, MAT requires to consume $4/3$ more power than TDMA.
Simulation results in Section \ref{evaluations} will confirm the observations highlighted in this section. In particular, it will be shown that SM-encoded MAT is expected to exhibit some performance benefits at low SNR over TDMA only for a per-user rate above 3 bit/s/Hz. 



\subsection{Diversity-Multiplexing Tradeoff (DMT)}
In this section, we derive the optimal DMT $\left(r,d^{\star}(r)\right)$, as defined in \cite{Zheng:2003}, of the MAT scheme and the ones achievable with several space-time encoded MAT architectures with QAM constellation. $r$ is the per-user multiplexing gain and $d^{\star}(r)$ is the optimal diversity gain at asymptotic high SNR. In \cite{Chen:2012}, an upper bound of the asymptotic DMT (at infinite SNR) for the MAT scheme was derived. We show in Theorem \ref{Theorem_MAT} that this upper bound is actually the exact DMT.
\begin{theorem}\label{Theorem_MAT} The asymptotic DMT $\left(r,d^{\star}(r)\right)$ of the MAT scheme over i.i.d.\ Rayleigh fading channel is the piecewise-linear function joining the points $(0,3)$, $(\frac{1}{3},1)$ and $(\frac{2}{3},0)$.
\end{theorem}
\par \textit{Proof:} The proof is provided in Appendix \ref{proof_DMT_MAT}.
$\hfill\Box$

Approximately universal codes, as defined in \cite{Tavildar:2006}, achieve the optimal DMT at infinite SNR for any fading distribution. The AMT strategy combined with such codes would therefore achieve the optimal SMT of Theorem \ref{Theorem_MAT}.

\begin{theorem} \label{Theorem_MAT_SM} The asymptotic DMT $\left(r,d(r)\right)$ achieved by SM-encoded MAT with ML decoding and QAM constellation over i.i.d.\ Rayleigh fading channel is given by $d(r)=2-3r$ for $r\in\left[0,\frac{2}{3}\right]$.
\end{theorem}
\par \textit{Proof:} The proof is provided in Appendix \ref{proof_DMT_SM_MAT_altMAT}.
$\hfill\Box$

We observe that a simple SM is suboptimal in the range $r\in\left[0,\frac{1}{3}\right]$.

\begin{theorem} \label{Theorem_MAT_OSTBC} The asymptotic DMT $\left(r,d(r)\right)$ achieved by O-STBC-encoded MAT with QAM constellation over i.i.d.\ Rayleigh fading channel is given by $d(r)=3\left(1-3r\right)$ for $r\in\left[0,\frac{1}{3}\right]$.
\end{theorem}

\par \textit{Proof:} Given the PEP expressions in \eqref{MAT_PEP_OSTBC_highSNR} at high SNR, the proof is straightforward and directly re-uses the derivations made for O-STBC in conventional Rayleigh fading MIMO channels \cite{Zheng:2003,Clerckx:2013}. 
$\hfill\Box$

O-STBC is clearly sub-optimal for any $r>0$. At $r=0$ (i.e.\ constant rate transmission), O-STBC achieves the maximum diversity gain (i.e.\ 3), which was confirmed from the PEP analysis.

\par For comparison, a simple TDMA transmission that ignores the delayed CSIT would achieve a DMT of $d(r)=2\left(1-2r\right)$ for a per-user multiplexing gain $r\in\left[0,\frac{1}{2}\right]$, i.e.\ the line joining the points $(0,2)$, $(\frac{1}{2},0)$. Transmission using O-STBC within a coherence time for each user at a time would achieve such DMT.  

\par It is important to recall that the DMT results are only valid at infinite SNR. While a characterization of the DMT at finite SNR is very challenging, a closer look at the proofs in the appendices highlights that the outage probability and the union bound on the error probability often scale as $\rho^{-m}\log\left(\rho^{-n}\right)$ (with $m>0$ and $n>0$) at high but finite SNR. While the $\log$ term does not impact the diversity gain at asymptotic high SNR, it does at finite SNR (similarly to the analysis made on the finite SNR average PEP). This explains why the PEP analysis highlights some benefits of TDMA at finite SNR even though the asymptotic DMT of TDMA is clearly lower than the optimal asymptotic DMT of MAT and that achieved with SM-encoded MAT with QAM constellation.

\subsection{Space-Time Code Design}\label{code_design}
From \eqref{MAT_PEP_highSNR}, focusing on the worst-case PEP, full rank codes should be designed in MAT such that 
\begin{equation}
d_{\lambda,MAT}=\max_{\mycom{\mathbf{C},\mathbf{E}}{\mathbf{C}\neq\mathbf{E}}}\left(\lambda_1 \lambda_2\right)^{-1}\frac{\log\left(\lambda_1\right)-\log\left(\lambda_2\right)}{\lambda_1-\lambda_2}
\end{equation}
is minimized, where $\lambda_1$ and $\lambda_2$ are the two eigenvalues of the error matrix $\tilde{\mathbf{E}}=\left(\mathbf{C}-\mathbf{E}\right)\left(\mathbf{C}-\mathbf{E}\right)^H$.

\par Compared to classical point-to-point i.i.d.\ Rayleigh fading channels, the coding gain is now a function of the quantity $\frac{\log\left(\lambda_1\right)-\log\left(\lambda_2\right)}{\lambda_1-\lambda_2}$.
We can write 
\begin{multline}
\max_{\mycom{\mathbf{C},\mathbf{E}}{\mathbf{C}\neq\mathbf{E}}}\frac{\log\left(\lambda_1\right)-\log\left(\lambda_2\right)}{\lambda_1-\lambda_2}\\
\geq \max_{\mycom{\mathbf{C},\mathbf{E}}{\mathbf{C}\neq\mathbf{E}}}\frac{2}{\lambda_1+\lambda_2} = \frac{2}{\min_{\mycom{\mathbf{C},\mathbf{E}}{\mathbf{C}\neq\mathbf{E}}}\left\|\mathbf{C}-\mathbf{E}\right\|_F^2}.\label{ineq_log}
\end{multline}
Equality occurs whenever the minimum coding gain error matrix (i.e.\ the one leading to the maximum of the left-hand side of \eqref{ineq_log}) has equal eigenvalues. If equality is achieved, and assuming we aim at minimizing the worst-case PEP, codes with large minimum trace among all error matrices should be favored.

\par Let us further assume a linear Space-Time Block Code whose codewords write as 
\begin{equation}\label{LDC_structure}	\mathbf{C}=\sum_{q=1}^{Q}\mathbf{\Phi}_{q}\Re{\left[c_q\right]}+\mathbf{\Phi}_{q+Q}\Im{\left[c_q\right]}
\end{equation}
with $\left\{\textnormal{Tr}\left\{\mathbf{\Phi}_{q}\mathbf{\Phi}_{q}^H\right\}=T/Q\right\}_{q=1}^{2Q}$ \cite{Clerckx:2013}. $\mathbf{\Phi}_q$ are complex basis matrices of size $n_t \times T$, $c_q$ stands for the complex information symbol, $\textit{Q}$ is the number of complex symbols $c_q$ transmitted over a codeword, $\Re$ and $\Im$ stand for the real and imaginary parts. From \cite{Clerckx:2013} (Proposition 6.4), denoting the minimum squared Euclidean distance of the constellation used by $d_{\textnormal{min}}^2$, we know that  
\begin{equation}\label{upperbound_min_trace}
\min_{\mycom{\mathbf{C},\mathbf{E}}{\mathbf{C}\neq\mathbf{E}}}\left\|\mathbf{C}-\mathbf{E}\right\|_F^2\leq \frac{T}{Q} d_{\textnormal{min}}^2
\end{equation}
and equality is achieved if the basis matrices $\left\{\mathbf{\Phi}_{q}\right\}_{q=1}^{2Q}$ satisfy the conditions 
\begin{equation}\label{eqn:LDC_4}
\textnormal{Tr}\left\{\mathbf{\Phi}_{q}\mathbf{\Phi}_{p}^{H}+\mathbf{\Phi}_{p}\mathbf{\Phi}_{q}^{H}\right\}=0, \hspace{0.3cm}  \mbox{$q\neq p$}
\end{equation}
or equivalently
\begin{equation}\label{eqn:LDC_7}
\mathcal{X}^T\mathcal{X}=\frac{T}{Q}\hspace{0.1cm}\mathbf{I}_{2Q}
\end{equation}
with \begin{equation}
\mathcal{X}=\left[\begin{array}{ccc}
\text{vec}\left(\left[
\begin{array}{l}
\Re\left[\mathbf{\Phi}_1\right]\\
\Im\left[\mathbf{\Phi}_1\right]
\end{array}\right]\right) & \cdots &
\text{vec}\left(\left[
\begin{array}{l}
\Re\left[\mathbf{\Phi}_{2Q}\right]\\
\Im\left[\mathbf{\Phi}_{2Q}\right]
\end{array}\right]\right)
\end{array}
\right].
\end{equation}

Combining \eqref{upperbound_min_trace} with \eqref{ineq_log}, we can write
\begin{equation}\label{lower_bound_LDC}
\max_{\mycom{\mathbf{C},\mathbf{E}}{\mathbf{C}\neq\mathbf{E}}}\frac{\log\left(\lambda_1\right)-\log\left(\lambda_2\right)}{\lambda_1-\lambda_2}\geq \frac{2 Q}{T d_{\textnormal{min}}^2}.
\end{equation}
From previous discussions, we conclude that the equality in \eqref{lower_bound_LDC} is achieved if the following two conditions are satisfied: 1) $\mathcal{X}^T\mathcal{X}=\frac{T}{Q}\hspace{0.1cm}\mathbf{I}_{2Q}$ 
and 2) the minimum coding gain error event (i.e.\ the one leading to the maximum of the left-hand side of \eqref{ineq_log}) is such that $\tilde{\mathbf{E}}=\frac{T d_{\textnormal{min}}^2}{2Q}\mathbf{I}_2$. We note that approximately universal codes as defined in \cite{Tavildar:2006} commonly satisfy condition 1 but do not satisfy condition 2 \cite{Clerckx:2013}. Combined with the classical min det maximization design criterion \cite{Tarokh:1998}, those two conditions provide further insights into how to enhance space-time code designs for BC with delayed CSIT. 

\subsection{A Larger Number of Users and Antennas}
Analysis and discussions have been limited to the two-user MAT so far. However the MAT strategy is also known for a general K-user scenario \cite{MAT:2012}. The extension of the error rate and DMT analysis to the K-user is beyond the scope of this paper. Nevertheless it is expected that the K-user MAT, similarly to the two-user scheme, will also be subject to a diversity ``loss'' (compared to classical Rayleigh distribution) at finite SNR owing to the double Rayleigh distribution of some of the channel coefficients.

\section{Spatially Correlated Fading Channels}\label{correlated_channel}
We now extend the discussion to spatially correlated channels with any transmit correlation matrices $\mathbf{R}_{t,1}$ and $\mathbf{R}_{t,2}$.
\subsection{Error Rate Performance}
Following the derivations in Appendix \ref{Appendix_MAT}, the average PEP of MAT is upper bounded as displayed in \eqref{MAT_PEP_correlated}
\begin{table*}
\begin{equation}\label{MAT_PEP_correlated}
P\left(\mathbf{C}\rightarrow \mathbf{E}\right)\leq\frac{1}{b_{1,1} b_{2,1}}\left[\frac{1}{b_{1,2} b_{2,2}}+\frac{\left(b_{2,2}-1\right)^2}{b_{2,2}^2\left(b_{1,2}-b_{2,2}\right)}\exp\left(\frac{1}{b_{2,2}}\right)\textnormal{Ei}\left(\frac{-1}{b_{2,2}}\right)-\frac{\left(b_{1,2}-1\right)^2}{b_{1,2}^2\left(b_{1,2}-b_{2,2}\right)}\exp\left(\frac{1}{b_{1,2}}\right)\textnormal{Ei}\left(\frac{-1}{b_{1,2}}\right)\right]
\end{equation}
\hrulefill
\end{table*}
where $b_{k,i}=1+a_{k,i}$ ($k,i=1,2$) with $a_{k,i}=\frac{\rho}{4}\lambda_{k,i}$ and $\lambda_{k,i}$ the $k^{th}$ eigenvalue of the matrix $\mathbf{R}_{t,i}\tilde{\mathbf{E}}$. 

Assuming a full rank code, i.e.\ $\lambda_{k,i}>0$ for $k,i=1,2$, the average PEP for large enough $a_{1,2}$ and $a_{2,2}$ can be approximated as
\begin{equation}
P\left(\mathbf{C}\rightarrow \mathbf{E}\right)\leq \left(b_{1,1} b_{2,1}\right)^{-1}\frac{\log\left(b_{1,2}\right)-\log\left(b_{2,2}\right)}{b_{1,2}-b_{2,2}}\label{MAT_PEP_mediumSNR_correlated}
\end{equation}
with $b_{1,1} b_{2,1}=\det\big(\mathbf{I}_2+\rho/4\mathbf{R}_{t,1}\tilde{\mathbf{E}}\big)$.

In the more restrictive condition that $a_{k,i}$ $\forall k,i$ are all large (i.e.\ high SNR),
\begin{align}
P\left(\mathbf{C}\rightarrow \mathbf{E}\right)&\approx \left(\frac{\rho}{4}\right)^{-3}\frac{1}{\lambda_{1,1} \lambda_{2,1}}\frac{\log\left(\lambda_{1,2}\right)-\log\left(\lambda_{2,2}\right)}{\lambda_{1,2}-\lambda_{2,2}}\label{MAT_PEP_highSNR_correlated}
\end{align}
with $\lambda_{1,1} \lambda_{2,1}=\det\left(\mathbf{R}_{t,1}\right)\det\big(\tilde{\mathbf{E}}\big)$.
The maximum achievable diversity gain is 3. User 1's performance is a function of user 2's spatial correlation matrix $\mathbf{R}_{t,2}$.


\par With SM-encoded MAT, the rank-1 error matrix $\tilde{\mathbf{E}}$ has a unique non-zero eigenvalue $\lambda_{1,i}$, $i=1,2$. Hence $\lambda_{2,i}=0$, $b_{2,i}=1$, $b_{1,i}=1+a_{1,i}$ and $b_{1,i}-b_{2,i}=a_{1,i}$ with $a_{1,i}=\frac{\rho}{4}\lambda_{1,i}$. The average PEP \eqref{MAT_PEP_correlated} simplifies into
\begin{equation}\label{MAT_PEP_SM_correlated}
P\left(\mathbf{C}\rightarrow \mathbf{E}\right)\leq \frac{1}{b_{1,1}}\left[\frac{1}{b_{1,2}}-\frac{a_{1,2}}{b_{1,2}^2}\exp\left(\frac{1}{b_{1,2}}\right)\textnormal{Ei}\left(\frac{-1}{b_{1,2}}\right)\right].
\end{equation}
For large $a_{1,2}$, $P\left(\mathbf{C}\rightarrow \mathbf{E}\right)\leq \frac{\log\left(b_{1,2}\right)}{b_{1,1} b_{1,2}}$, leading to
\begin{multline}\label{MAT_PEP_SM_mediumSNR_correlated}
P\left(\mathbf{C}\rightarrow \mathbf{E}\right)\leq \log\left(1+\frac{\rho}{4}\textnormal{Tr}\big\{\mathbf{R}_{t,2}\tilde{\mathbf{E}}\big\}\right)\\\left(1+\frac{\rho}{4}\textnormal{Tr}\big\{\mathbf{R}_{t,e}\tilde{\mathbf{E}}\big\}+\left(\frac{\rho}{4}\right)^2\textnormal{Tr}\big\{\mathbf{R}_{t,1}\tilde{\mathbf{E}}\big\}\textnormal{Tr}\big\{\mathbf{R}_{t,2}\tilde{\mathbf{E}}\big\}\right)^{-1}
\end{multline}
where $\mathbf{R}_{t,e}=\mathbf{R}_{t,1}+\mathbf{R}_{t,2}$. 
Bound \eqref{MAT_PEP_SM_mediumSNR_correlated} should be contrasted with the performance of conventional MU-MIMO and point-to-point MIMO. This will be discussed in detail in Section \ref{Discussions}.
 
\par With O-STBC (Alamouti code) encoded MAT, $\tilde{\mathbf{E}}=\alpha \mathbf{I}_2$, $b_{k,i}=1+\frac{\rho}{4}\alpha\lambda_{k,i}$. Assuming $a_{1,2}$ and $a_{2,2}$ are large enough, the average PEP in \eqref{MAT_PEP_mediumSNR_correlated} simplifies as
\begin{multline}
P\left(\mathbf{C}\rightarrow \mathbf{E}\right)\leq \left(\det\left(\mathbf{I}_2+\frac{\rho}{4}\alpha\mathbf{R}_{t,1}\right)\right)^{-1}\\\frac{\log\left(1+\frac{\rho}{4}\alpha\lambda_{1,2}\right)-\log\left(1+\frac{\rho}{4}\alpha\lambda_{2,2}\right)}{\frac{\rho}{4}\alpha\left(\lambda_{1,2}-\lambda_{2,2}\right)}.\label{MAT_PEP_mediumSNR_correlated_OSTBC}
\end{multline}
At high SNR, 
\begin{align}
P\left(\mathbf{C}\rightarrow \mathbf{E}\right)
&\leq \left(\frac{\rho}{4}\right)^{-3}\alpha^{-3} \left(\det\left(\mathbf{R}_{t,1}\right)\right)^{-1}\nonumber\\
&\hspace{1cm}\frac{\log\left(\lambda_{1}\left(\mathbf{R}_{t,2}\right)\right)-\log\left(\lambda_{2}\left(\mathbf{R}_{t,2}\right)\right)}{\lambda_{1}\left(\mathbf{R}_{t,2}\right)-\lambda_{2}\left(\mathbf{R}_{t,2}\right)},\nonumber\\
&= \left(\frac{\rho}{4}\right)^{-3}\alpha^{-3} \left(1-\left|t_1\right|^2\right)^{-1}\nonumber\\
&\hspace{1cm}\frac{\log\left(1+\left|t_2\right|\right)-\log\left(1-\left|t_2\right|\right)}{2\left|t_2\right|}.
\label{MAT_PEP_highSNR_correlated_OSTBC}
\end{align}
The higher the magnitude of the correlation coefficients $\left|t_1\right|$ and $\left|t_2\right|$, the higher the error rate.

\subsection{MISO BC with Outdated CSIT vs.\ MU-MIMO vs.\ Point-to-Point MIMO}\label{Discussions}

We can now make interesting observations about the PEP behaviour of \eqref{MAT_PEP_SM_mediumSNR_correlated} and contrasts with conventional MU-MIMO with perfect CSIT and point-to-point MIMO (single-user SM and TDMA based on O-STBC): 
\begin{itemize} 
\item \textit{Subspace alignment:} $\textnormal{Tr}\big\{\mathbf{R}_{t,x}\tilde{\mathbf{E}}\big\}$ (with $x=e,1,2$) suggests that the performance at low/medium SNR and high SNR depends on the alignment between the eigenvectors of the error matrix $\tilde{\mathbf{E}}$ and those of the correlation matrix $\mathbf{R}_{t,x}$ (i.e.\ $\mathbf{R}_{t,e}$, $\mathbf{R}_{t,1}$ and $\mathbf{R}_{t,2}$), which is reminiscent of the point-to-point MIMO behaviour with Spatial Multiplexing \cite{Clerckx:2007b}.
The error probability is primarily a function of the worst-case alignment, which is known to be detrimental to Spatial Multiplexing performance \cite{Clerckx:2007,Clerckx:2007b}. Similarly such subspace alignment is expected to be detrimental to the performance of SM-encoded MAT. 

\item \textit{User orthogonality:} $\textnormal{Tr}\big\{\mathbf{R}_{t,e}\tilde{\mathbf{E}}\big\}$ is not a function of each individual transmit correlation matrix but only of their sum $\mathbf{R}_{t,e}$. As $\mathbf{R}_{t,e}$ acts as an effective transmit correlation matrix, $\mathbf{R}_{t,e}$ can appear close to an identity matrix even though each user experiences highly transmit correlated channels. Assume for instance that user 1 and user 2 experience the same magnitude of transmit correlation but different phase, i.e.\ $t_1=|t|e^{j\varphi_1}$ and $t_2=|t|e^{j\varphi_2}$. For statistically orthogonal users with $\varphi_2-\varphi_1=\pi$, $\mathbf{R}_{t,e}=2 \mathbf{I}_2$ $\forall |t|$. Hence, despite the presence of transmit correlated channels, each user experiences the same performance as if the channels were independent ($t_1=t_2=0$). This shows that as the phase shift $\varphi_2-\varphi_1$ increases and users get more and more statistically orthogonal to each other ($\varphi_2-\varphi_1=\pi$), the performance is enhanced. This behaviour is reminiscent of conventional MU-MIMO \cite{Clerckx:2013,Gesbert:2007,Clerckx:2008}. 

\item \textit{Transmit correlation:}
Transmit correlation is known to be detrimental to point-to-point MIMO (using SM and O-STBC) with no channel state/distribution knowledge at the transmitter  \cite{Clerckx:2007,Clerckx:2007b}. The analysis of SM-encoded MAT highlights that the performance in spatially correlated channels approaches that of uncorrelated channels as long as $|t_1|\approx|t_2|$ and $\phi\approx \pi$ (i.e.\ user channels are statistically orthogonal), so that $\mathbf{R}_{t,e}$ effectively behaves as if transmit correlation is zero. Otherwise, transmit correlation would be detrimental to the performance. In particular, asymmetric scenarios where one of the correlation coefficients is high and the other one low ($|t_1|>>|t_2|$ or inversely) would lead to a higher error rate than if both coefficients were large and $\phi\approx \pi$. Hence, combined with a suitable user pairing, SM-encoded MAT is expected to be less sensitive to spatial correlation than TDMA.
\end{itemize}

\par It is interesting to make an analogy between $\mathbf{R}_{t,e}$ and the transmit correlation matrix of a point-to-point MIMO channel, denoted as $\mathbf{R}_t$. Indeed, $\mathbf{R}_{t}$ can always be decomposed into the sum of two matrices $\mathbf{R}_{t,A}$ and $\mathbf{R}_{t,B}$ so that $\mathbf{R}_{t}=\mathbf{R}_{t,A}+\mathbf{R}_{t,B}$. This originates from decomposing all channel multipaths into two independent clusters, denoted as $A$ and $B$. $\mathbf{R}_{t,x}$, with $x=A,B$, can be viewed as the transmit correlation matrix accounting for the cluster $x$ of multipaths. Choosing statistically orthogonal users so as to decrease the off-diagonal entries of $\mathbf{R}_{t,e}$ can therefore be viewed as an hypotetical point-to-point MIMO channel where we would have the flexibility to adjust the clusters $A$ and $B$ of multipaths so that the transmit correlation matrix $\mathbf{R}_{t}$ is better conditioned. 

\par Note that \cite{Wang:2013} also pointed out the presence of the quantity $\mathbf{R}_{t,e}$ in their analysis. However it originated from the use of Jensen's inequality and was therefore only observed in an upper bound of the achievable rate. The analysis here reflects the true presence of such quantity in the PEP.

\subsection{User Pairing with Outdated CSIT}\label{user_pairing_section}
Observations made in previous section provide useful guidelines to appropriately pair users with outdated CSIT. Let us assume a SM-encoded MAT for simplicity. The user pairing strategy should be SNR dependent.
\par At \textit{low to medium SNR}, statistically orthogonal users (characterized by $\varphi_2-\varphi_1\approx\pm \pi$) with $\left|t_1\right|\approx\left|t_2\right|$ should be paired together so as to experience an identity effective transmit correlation matrix $\mathbf{R}_{t,e}$. 
\par At \textit{high SNR}, on the other hand, users should be paired so as to maximize $\textnormal{Tr}\big\{\mathbf{R}_{t,1}\tilde{\mathbf{E}}\big\}\textnormal{Tr}\big\{\mathbf{R}_{t,2}\tilde{\mathbf{E}}\big\}$ over all possible error matrices, i.e.\ users should be paired such that the dominant eigenvector of each error matrix is never aligned with the weakest eigenvector of either $\mathbf{R}_{t,1}$ or $\mathbf{R}_{t,2}$. Pairing statistically orthogonal users helps decreasing the error rate as confirmed by the simulations in Section \ref{evaluations}.



\subsection{Robust Code Design}
Similarly to robust code design in point-to-point MIMO channels where space-time codes and/or precoders are designed so as to avoid the detrimental alignment between eigenvectors of the error matrix and the transmit correlation matrix, precoder/code could be made robust for MISO BC with outdated CSIT. The derivations above combined with those made in \cite{Clerckx:2007,Clerckx:2007b,Clerckx:2008c} would provide useful guidelines to derive such precoders/codes. Nevertheless, given the similarities between MISO BC with outdated CSIT and point-to-point MIMO channels, it is expected that robust precoders designed for SM would be suitable candidates for SM-encoded MAT.

\subsection{Signal Constellation Optimization}

Assuming the transmit correlation matrices are known to the transmitter, we resort to an optimization of non-linear signal constellations in order to improve the performance of SM-encoded MAT. The design relies on an extension of the results obtained in \cite{Clerckx:2004} to the MAT transmission. In the non-linear signal constellations, the first 
entry of the $2 \times 1$ codeword $\mathbf{C}$ (assume $T=1$) is selected from a constellation $\mathbf{S}$ (containing the
symbols $S_m$, with $m=1,\ldots,M_0$) and the second entry of $\mathbf{C}$ is
selected from a constellation $\mathbf{Q}_m$ (containing the symbols $Q_{mn}$,
with $n=1,\ldots,M_1$), which depends on the symbol $S_m$ chosen as the first entry of the codeword. Therefore, the second entry of $\mathbf{C}$ is no longer independent from the first entry. The codewords write as $\mathbf{C}=\left[\begin{array}{cc}S_m & Q_{mn} \end{array}\right]^T$. Contrary to the point-to-point MIMO, the dependence between $\mathbf{S}$ and $\mathbf{Q}_m$ is now a function of two transmit correlation matrices rather than one. For i.i.d.\ channels, $\mathbf{S}$ and $\mathbf{Q}_m$ simply boil down to classical QAM constellations.

\par The average PEP of SM-encoded MAT in spatially correlated channels shows that the performance is highly determined by the quantities $\textnormal{Tr}\big\{\mathbf{R}_{t,1}\tilde{\mathbf{E}}\big\}$ and $\textnormal{Tr}\big\{\mathbf{R}_{t,2}\tilde{\mathbf{E}}\big\}$. In particular, from \eqref{MAT_PEP_SM_mediumSNR_correlated}, both users' performance in MAT are function of the quantity $\big(1+\frac{\rho}{4}\textnormal{Tr}\big\{\mathbf{R}_{t,1}\tilde{\mathbf{E}}\big\}\big)\big(1+\frac{\rho}{4}\textnormal{Tr}\big\{\mathbf{R}_{t,2}\tilde{\mathbf{E}}\big\}\big)$. In order to design the nonlinear constellations, we introduce the following objective function
\begin{multline}\label{average_SER}
\bar{P}\approx\frac{1}{M_0 M_1}\sum_{m=1}^{M_0}\sum_{u=1}^{M_0}\sum_{n=1}^{M_1}\sum_{v=1}^{M_1}
s\left(S_m,S_u,Q_{mn},Q_{uv}\right)\\
\prod_{i=1}^{2}\left(1+\frac{\rho}{4}\textnormal{Tr}\big\{\mathbf{R}_{t,i}\tilde{\mathbf{E}}\big\}\right)^{-1},
\end{multline}
with 
\begin{multline}
\textnormal{Tr}\big\{\mathbf{R}_{t,i}\tilde{\mathbf{E}}\big\}=\left|S_m-S_u\right|^2+\left|Q_{mn}-Q_{uv}\right|^2\\
+2\Re\left\{t_i\left(S_m-S_u\right)\left(Q_{mn}-Q_{uv}\right)^*\right\}.
\end{multline}
The weights, denoted as $s$, result from the fact that different codeword vectors $\mathbf{C}$ and $\mathbf{E}$ may cause a different number of symbol errors, i.e.,\
\begin{equation}
s\left(c_0,e_0,c_1,e_1\right)=\left\{\begin{array}{l}
2, \hspace{0.2cm} c_0-e_0\neq 0 \hspace{0.2cm}\text{and}\hspace{0.2cm} c_1-e_1\neq 0, \\
1, \hspace{0.2cm} c_0-e_0=0 \hspace{0.2cm} \text{or}\hspace{0.2cm} c_1-e_1=0, \\
0, \hspace{0.2cm} c_0-e_0=0 \hspace{0.2cm} \text{and}\hspace{0.2cm} c_1-e_1=0. 
\end{array}
\right.
\end{equation}
$\bar{P}$ can be seen as an estimate of the average symbol error probability (whose accuracy was demonstrated in \cite{Clerckx:2004} and references therein) accounting only for the term that is common to both users' performance in MAT. 
The problem is the selection of the signal constellations $\mathbf{S}$ and $\mathbf{Q}_m$ to minimize $\bar{P}$ under an average power constraint \cite{Clerckx:2004}. A constrained gradient-search algorithm is used to determine the optimum constellations. Let $\mathbf{F}$ be the vector defined as
\begin{equation}
\mathbf{F}=\left[\begin{array}{cccc}S_1 \ldots S_{M_0} & Q_{11} \ldots Q_{1 M_1} & \ldots & Q_{M_0 1} \ldots Q_{M_0 M_1}\end{array}\right]^T
\end{equation}
and $\mathbf{F}^k$ the vector $\mathbf{F}$ at the $k^{th}$ step of the algorithm. The unconstrained
gradient algorithm is described by
\begin{equation}\label{constrained_gradient}
\mathbf{F}^{k+1}=\mathbf{F}^k-\alpha \nabla \bar{P}\left(\mathbf{F}^k\right)
\end{equation}
where $\alpha$ is the step size and $\bar{P}\left(\mathbf{F}^k\right)$ is the gradient of $\bar{P}$ with respect to $\mathbf{F}^k$. The constellations are normalized at every iteration in order to account for the power constraint. The expressions of the gradients are provided in Appendix \ref{Appendix_gradient}.


\section{Other Strategies Relying on Delayed CSIT}\label{Section_Alt_MAT}
Aside MAT, other interesting strategies have been proposed to exploit delayed CSIT \cite{Yang:2013,Chen:2012}. Alternative MAT strategy is quite appealing as it has been shown to be a useful building block of a larger scheme \cite{Yang:2013}. Moreover it remains quite tractable and spans 3 channel uses similarly to MAT. The scheme proposed in \cite{Chen:2012} has the benefit of achieving a diversity gain of 6 (at infinite SNR), hence higher than that of MAT and Alternative MAT, but spans 24 channel uses. Thanks to the higher diversity gain, it is expected to have a better PEP performance than MAT/Alternative MAT in the infinite SNR regime but would also be subject to the effects due to the double Rayleigh distribution of some channel coefficients at finite SNR (similarly to MAT and Alternative MAT). In the sequel, we extend the past discussions to the Alternative MAT.

\subsection{Space-Time Encoded Alternative MAT}
\par Alternative MAT performs transmission in a different manner than MAT. On time slot $t$, Alternative MAT transmits $\mathbf{x}_{1,t}=\sqrt{\rho}\left(\mathbf{c}_{t}+\mathbf{c}'_{t}\right)$ in coherence time 1, user 1's overheard interference $\mathbf{x}_{2,t}=\sqrt{\rho}\left[\begin{array}{cc} \mathbf{h}_1\mathbf{c}'_{t} & 0 \end{array}\right]^T$ in coherence time 2 and user 2's overheard interference $\mathbf{x}_{3,t}=\sqrt{\rho}\left[\begin{array}{cc} \mathbf{g}_1\mathbf{c}_{t} & 0 \end{array}\right]^T$ in coherence time 3 \cite{Yang:2013}. Hence, for a space-time encoded Alternative MAT, the equivalent system model for user 1 at time instant $t=1,\ldots,T$ can be written as
\begin{multline}
\left[\begin{array}{c}y_{1,t}\\y_{2,t}\\y_{3,t}\end{array}\right]=\sqrt{\rho}\left[\begin{array}{cc} h_{1,1} & h_{1,2} \\ 0 & 0 \\ h_{3,1}g_{1,1} & h_{3,1} g_{1,2}\end{array}\right]\mathbf{c}_{t}\\+\sqrt{\rho}\left[\begin{array}{cc} h_{1,1} & h_{1,2} \\ h_{2,1} h_{1,1} & h_{2,1} h_{1,2} \\ 0 & 0\end{array}\right]\mathbf{c}'_{t}+\left[\begin{array}{c}n_{1,t}\\n_{2,t}\\n_{3,t}\end{array}\right], 
\end{multline}
and after further interference elimination
\begin{equation}\label{system_model_AltMAT}
\tilde{\mathbf{y}}_{t}=\left[\begin{array}{c}y_{3,t}\\h_{2,1} y_{1,t}-y_{2,t}\end{array}\right]=\sqrt{\rho}\mathbf{H}\mathbf{c}_{t}+\left[\begin{array}{c}n_{3,t}\\ h_{2,1} n_{1,t}-n_{2,t}\end{array}\right], 
\end{equation}
where 
\begin{equation}\label{H_system_model_AltMAT}
\mathbf{H}=\left[\begin{array}{cc} h_{3,1}g_{1,1} & h_{3,1} g_{1,2} \\h_{2,1}h_{1,1} & h_{2,1}h_{1,2} \end{array}\right].
\end{equation}
ML decoder as in \eqref{equation2} is performed where $\mathbf{\Sigma}=\textnormal{diag}\big\{1,1+\left|h_{2,1}\right|^2\big\}$ is the covariance matrix of the noise vector in \eqref{system_model_AltMAT}. The minimization in \eqref{equation2} is performed over all possible codeword vectors $\mathbf{C}$. 

For Alternative MAT, we can define
\begin{equation}
\tilde{\mathbf{H}}=\mathbf{\Sigma}^{-1/2}\mathbf{H}=\left[\begin{array}{cc}Z & 0 \\ 0 & X\end{array}\right]\underbrace{\left[\begin{array}{cc} g_{1,1} & g_{1,2} \\ h_{1,1} & h_{1,2}\end{array}\right]}_{\mathbf{H}'}
\end{equation}
where $Z=h_{3,1}$ and $X=\frac{h_{2,1}}{\sqrt{1+\left|h_{2,1}\right|^2}}$, such that the conditional PEP can be written as \eqref{MAT_PEP_cond}. Contrary to MAT, the average PEP of Alternative MAT is obtained by taking the expectation of the conditional PEP over the distribution of $\mathbf{H}'$ and then over $X$ and $Z$.

\subsection{Error Rate Performance of (encoded) Alternative MAT}
Similar derivations can be made for the Alternative MAT. Let us assume first I.I.D.\ Rayleigh Fading Channels.
Following results derived in Appendix \ref{Appendix_Alt_MAT}, the average PEP can be written as displayed in \eqref{Alt_MAT_PEP}.
\begin{table*}
\begin{equation}\label{Alt_MAT_PEP}
P\left(\mathbf{C}\rightarrow \mathbf{E}\right)\leq\left[\frac{\exp\left(\frac{1}{a_2}\right)\textnormal{Ei}\left(\frac{-1}{a_2}\right)-\exp\left(\frac{1}{a_1}\right)\textnormal{Ei}\left(\frac{-1}{a_1}\right)}{a_1-a_2}\right]\left[\frac{1}{b_1 b_2}+\frac{\left(b_2-1\right)^2}{b_2^2\left(b_1-b_2\right)}\exp\left(\frac{1}{b_2}\right)\textnormal{Ei}\left(\frac{-1}{b_2}\right)-\frac{\left(b_1-1\right)^2}{b_1^2\left(b_1-b_2\right)}\exp\left(\frac{1}{b_1}\right)\textnormal{Ei}\left(\frac{-1}{b_1}\right)\right]
\end{equation}
\hrulefill
\end{table*}
At high SNR, assuming full rank code, the average PEP \eqref{Alt_MAT_PEP} can be approximated as
\begin{align}
P\left(\mathbf{C}\rightarrow \mathbf{E}\right)&\leq \left(\frac{\log\left(a_1\right)-\log\left(a_2\right)}{a_1-a_2}\right)^2\nonumber\\
&=\left(\frac{\rho}{4}\right)^{-2}\left(\frac{\log\left(\lambda_1\right)-\log\left(\lambda_2\right)}{\lambda_1-\lambda_2}\right)^2.\label{Alt_MAT_PEP_highSNR}
\end{align}
The maximum achievable diversity gain is 2 and the coding gain is proportional to $\frac{\log\left(\lambda_1\right)-\log\left(\lambda_2\right)}{\lambda_1-\lambda_2}$. 
Comparing \eqref{MAT_PEP_highSNR} and \eqref{Alt_MAT_PEP_highSNR}, we note the loss incurred by the Alternative MAT compared to the original MAT in terms of diversity and coding gains. The error probability of Alternative MAT is therefore expected to be significantly higher than that of MAT. Intuitively, a maximum diversity gain of 2 (rather than 3 or 4) is achieved because of the presence of $h_{3,1}$ and $h_{2,1}$ in both entries of respectively the first and second rows of $\mathbf{H}$ in \eqref{H_system_model_AltMAT}. This implies that an error is likely to occur whenever the two channel coefficients $h_{3,1}$ and $h_{2,1}$ are in deep fade.

\par From \eqref{Alt_MAT_PEP_highSNR}, we also note that if we aim at minimizing the worst-case PEP, 
\begin{equation}
d_{\lambda,Alt MAT}=\max_{\mycom{\mathbf{C},\mathbf{E}}{\mathbf{C}\neq\mathbf{E}}}\left(\frac{\log\left(\lambda_1\right)-\log\left(\lambda_2\right)}{\lambda_1-\lambda_2}\right)^2
\end{equation}
should be minimized. Hence the code design discussed in Section \ref{code_design} also applies to Alternative MAT.

\par For SM-encoded Alternative MAT, the average PEP becomes
\begin{multline}\label{Alt_MAT_PEP_SM}
P\left(\mathbf{C}\rightarrow \mathbf{E}\right)\leq \frac{1}{a}\exp\left(\frac{1}{a}\right)\textnormal{Ei}\left(\frac{-1}{a}\right)\\
\left[\frac{a}{\left(1+a\right)^2}\exp\left(\frac{1}{1+a}\right)\textnormal{Ei}\left(\frac{-1}{1+a}\right)-\frac{1}{1+a}\right].
\end{multline}
At high SNR,
\begin{equation}\label{SM_highSNR_PEP_AltMAT}
P\left(\mathbf{C}\rightarrow \mathbf{E}\right)\leq \left(\frac{\log\left(a\right)}{a}\right)^2=\left(\frac{\rho}{4}\right)^{-2}\left(\log\left(\frac{\rho}{4}\lambda\right)\right)^2.
\end{equation}
The diversity gain at high SNR writes as $d=2\left(1-\frac{1}{\log\left(a\right)}\right)\approx 2$. Due to the $(\log(\rho))^2$ term, following the discussion in Section \ref{iid_error_rate_section}, the SNR at which Alternative MAT is outperformed by TDMA is even smaller than that of MAT. 


\par With O-STBC-encoded Alternative MAT, the average PEP is written as
\begin{multline}\label{Alt_MAT_PEP_OSTBC}
P\left(\mathbf{C}\rightarrow \mathbf{E}\right)\leq \frac{1}{a^2}\left[a+\exp\left(\frac{1}{a}\right)\textnormal{Ei}\left(\frac{-1}{a}\right)\right]\\ \left[\frac{b^3-b^2+b}{b^4}+\frac{\left(1-b\right)^2}{b^4}\exp\left(\frac{1}{b}\right)\textnormal{Ei}\left(\frac{-1}{b}\right)\right],
\end{multline}
which leads at high SNR to
\begin{align}\label{Alt_MAT_PEP_OSTBC_highSNR}
P\left(\mathbf{C}\rightarrow \mathbf{E}\right)&\leq \left[\frac{1}{a}-\frac{\log\left(a\right)}{a^2}\right]\left[\frac{1}{a}+\frac{\log\left(a\right)}{a^2}\right]\nonumber\\
&=\frac{1}{a^2}-\left(\frac{\log\left(a\right)}{a^2}\right)^2\approx\frac{1}{a^2}.
\end{align}
Similarly to SM-encoded Alternative MAT, a diversity gain of 2 is also achieved with O-STBC.

The performance of Alternative MAT can also be evaluated in spatially correlated Rayleigh fading channels. 
Following Appendix \ref{Appendix_Alt_MAT}, the average PEP can be written as displayed in \eqref{Alt_MAT_PEP_correlated}.
\begin{table*}
\begin{multline}\label{Alt_MAT_PEP_correlated}
P\left(\mathbf{C}\rightarrow \mathbf{E}\right)\leq\left[\frac{\exp\left(\frac{1}{a_{2,2}}\right)\textnormal{Ei}\left(\frac{-1}{a_{2,2}}\right)-\exp\left(\frac{1}{a_{1,2}}\right)\textnormal{Ei}\left(\frac{-1}{a_{1,2}}\right)}{a_{1,2}-a_{2,2}}\right]\\\left[\frac{1}{b_{1,1} b_{2,1}}+\frac{\left(b_{2,1}-1\right)^2}{b_{2,1}^2\left(b_{1,1}-b_{2,1}\right)}\exp\left(\frac{1}{b_{2,1}}\right)\textnormal{Ei}\left(\frac{-1}{b_{2,1}}\right)-\frac{\left(b_{1,1}-1\right)^2}{b_{1,1}^2\left(b_{1,1}-b_{2,1}\right)}\exp\left(\frac{1}{b_{1,1}}\right)\textnormal{Ei}\left(\frac{-1}{b_{1,1}}\right)\right]
\end{multline}
\hrulefill
\end{table*}
At high SNR, assuming full rank code, the average PEP \eqref{Alt_MAT_PEP_correlated} can be approximated as
\begin{equation}
P\left(\mathbf{C}\rightarrow \mathbf{E}\right)\leq \left(\frac{\rho}{4}\right)^{-2}\prod_{i=1}^2\frac{\log\left(\lambda_{1,i}\right)-\log\left(\lambda_{2,i}\right)}{\lambda_{1,i}-\lambda_{2,i}}.\label{Alt_MAT_PEP_correlated_highSNR}
\end{equation}

\par For SM, the average PEP becomes
\begin{multline}\label{Alt_MAT_PEP_SM_correlated}
P\left(\mathbf{C}\rightarrow \mathbf{E}\right)\leq \frac{1}{a_{1,2}}\exp\left(\frac{1}{a_{1,2}}\right)\textnormal{Ei}\left(\frac{-1}{a_{1,2}}\right)\\\left[\frac{a_{1,1}}{b_{1,1}^2}\exp\left(\frac{1}{b_{1,1}}\right)\textnormal{Ei}\left(\frac{-1}{b_{1,1}}\right)-\frac{1}{b_{1,1}}\right].
\end{multline}
At high SNR,
\begin{align}
P\left(\mathbf{C}\rightarrow \mathbf{E}\right)&\leq \left(\frac{\rho}{4}\right)^{-2} \frac{\log\left(\frac{\rho}{4}\lambda_{1,1}\right)}{\lambda_{1,1}}\frac{\log\left(\frac{\rho}{4}\lambda_{1,2}\right)}{\lambda_{1,2}},\nonumber\\
&=\left(\frac{\rho}{4}\right)^{-2}\frac{\log\left(\frac{\rho}{4}\textnormal{Tr}\big\{\mathbf{R}_{t,1}\tilde{\mathbf{E}}\big\}\right)\log\left(\frac{\rho}{4}\textnormal{Tr}\big\{\mathbf{R}_{t,2}\tilde{\mathbf{E}}\big\}\right)}{\textnormal{Tr}\big\{\mathbf{R}_{t,1}\tilde{\mathbf{E}}\big\}\textnormal{Tr}\big\{\mathbf{R}_{t,2}\tilde{\mathbf{E}}\big\}}.\label{AltMAT_SM_highSNR_correlated}
\end{align}

\par With O-STBC (Alamouti code), the average PEP \eqref{Alt_MAT_PEP_correlated_highSNR} at high SNR is written as
\begin{align}
P\left(\mathbf{C}\rightarrow \mathbf{E}\right)&\leq \left(\frac{\rho}{4}\right)^{-2}\alpha^{-2} \prod_{i=1}^2\frac{\log\left(\lambda_{1}\left(\mathbf{R}_{t,i}\right)/\lambda_{2}\left(\mathbf{R}_{t,i}\right)\right)}{\lambda_{1}\left(\mathbf{R}_{t,i}\right)-\lambda_{2}\left(\mathbf{R}_{t,i}\right)},\nonumber\\
&= \left(\frac{\rho}{4}\right)^{-2}\alpha^{-2} \prod_{i=1}^2\frac{\log\left(\left(1+\left|t_i\right|\right)/\left(1-\left|t_i\right|\right)\right)}{2\left|t_i\right|}.
\end{align}
Here also, similarly to MAT with O-STBC, the transmit correlation leads to a degradation of the PEP performance.

\subsection{Diversity-Multiplexing Tradeoff (DMT)}
We derive the asymptotic DMT of Alternative MAT in I.I.D.\ Rayleigh Fading Channels.
\begin{theorem} \label{Theorem_AltMAT} The asymptotic DMT $\left(r,d^{\star}(r)\right)$ of the Alternative MAT scheme over i.i.d.\ Rayleigh fading channel is given by $d^{\star}(r)=2-3r$ for $r\in\left[0,\frac{2}{3}\right]$, i.e.\ the piecewise-linear function joining the points $(0,2)$, $(\frac{1}{3},1)$ and $(\frac{2}{3},0)$.
\end{theorem}
\par \textit{Proof:} The proof is similar to that of Theorem \ref{Theorem_MAT} and is therefore omitted for brevity.
$\hfill\Box$

Comparing Theorem \ref{Theorem_MAT} and \ref{Theorem_AltMAT}, we observe that the asymptotic DMT of Alternative MAT is lower than that of MAT in the region $r\in\left[0,\frac{1}{3}\right]$.

\begin{theorem} \label{Theorem_AltMAT_SM} The asymptotic DMT $\left(r,d(r)\right)$ achieved by SM-encoded Alternative MAT with ML decoding and QAM constellation over i.i.d.\ Rayleigh fading channel is given by $d(r)=2-3r$ for $r\in\left[0,\frac{2}{3}\right]$.
\end{theorem}
\par \textit{Proof:} The proof is provided in Appendix \ref{proof_DMT_SM_MAT_altMAT}.
$\hfill\Box$

The DMT achieved by SM in Theorem \ref{Theorem_AltMAT_SM} is the same as that in Theorem \ref{Theorem_MAT_SM}.
We observe that a simple SM (i.e.\ no spatial encoding) is sufficient to achieve the optimal DMT with Alternative MAT but is suboptimal in the range $r\in\left[0,\frac{1}{3}\right]$ with MAT.

\begin{theorem} \label{Theorem_AltMAT_OSTBC} The asymptotic DMT $\left(r,d(r)\right)$ achieved by O-STBC-encoded Alternative MAT with QAM constellation over i.i.d.\ Rayleigh fading channel is given by $d(r)=2\left(1-3r\right)$ for $r\in\left[0,\frac{1}{3}\right]$.
\end{theorem}

\par \textit{Proof:} Given the PEP expressions in \eqref{Alt_MAT_PEP_OSTBC_highSNR} at high SNR, the proof is straightforward and directly re-uses the derivations made for O-STBC in conventional Rayleigh fading MIMO channels \cite{Zheng:2003,Clerckx:2013}. 
$\hfill\Box$

\begin{figure}
\centerline{\includegraphics[width=\columnwidth]{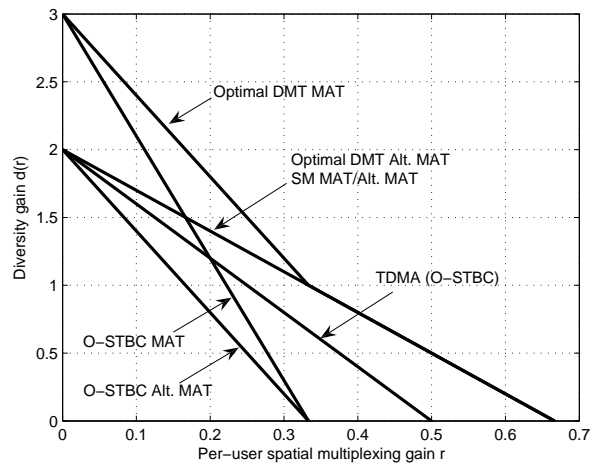}}
  \caption{Optimal DMT of MAT and Alternative MAT and achievable DMT with SM and O-STBC encoded MAT/Alternative MAT and TDMA.}
  \label{DMT_figure}
\end{figure}

The DMT achieved in Theorem \ref{Theorem_AltMAT_OSTBC} is lower than that obtained in Theorem \ref{Theorem_MAT_OSTBC}. 
Figure \ref{DMT_figure} summarizes all DMT at infinite SNR derived in Theorem \ref{Theorem_MAT} to \ref{Theorem_AltMAT_OSTBC}.

\section{Performance Evaluations}\label{evaluations}

%

\par
\begin{figure}
\centerline{\includegraphics[width=\columnwidth]{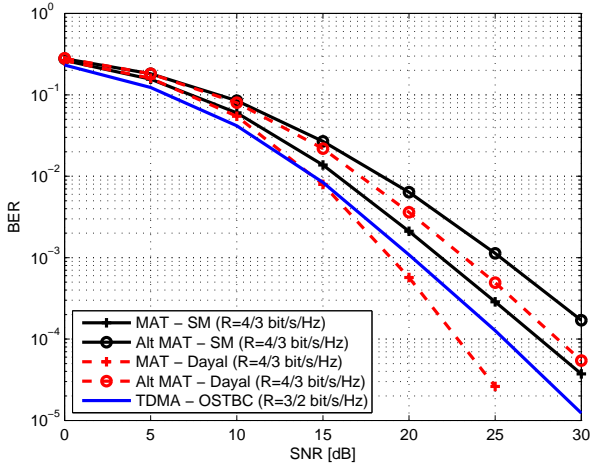}}
  \caption{Average Bit Error Rate (BER) of SM and ``Dayal'' encoded MAT and Alternative MAT with a per-user rate R=4/3 bit/s/Hz (corresponding to using QPSK) and comparison with TDMA with a per-user rate R=3/2 bit/s/Hz (O-STBC with 8PSK).}
  \label{BER_figure}
\end{figure}

\begin{figure}
\centerline{\includegraphics[width=\columnwidth]{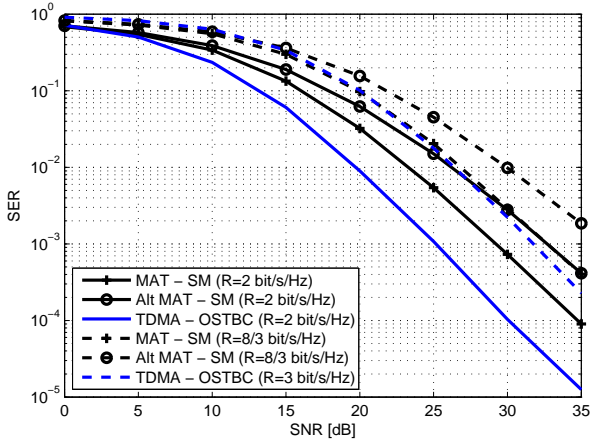}}
  \caption{Average Symbol Error Rate (SER) of SM-encoded MAT and Alternative MAT with a per-user rate R=2 bit/s/Hz (corresponding to using 8PSK) and R=8/3 bit/s/Hz (corresponding to using 16-QAM) and comparison with TDMA with a per-user rate R=2 bit/s/Hz (O-STBC with 16-QAM) and R=3 bit/s/Hz (O-STBC with 64-QAM).}
  \label{BER_figure_2}
\end{figure}

 Figure \ref{BER_figure} illustrates the BER performance of MAT and Alternative MAT with SM and a full-rate full-diversity approximately universal space-time code, denoted as ``Dayal'' \cite{Dayal:2005} over i.i.d.\ Rayleigh fading channels. QPSK is assumed so that it corresponds to a 4/3-bit/s/Hz transmission per user with ``MAT-SM'', ``Alt MAT-SM'', ``MAT-Dayal'', ``Alt MAT-Dayal''. As a baseline, we also display the performance of TDMA based on O-STBC and 8PSK over a conventional point-to-point MISO i.i.d.\ Rayleigh slow fading channels with two transmit antennas, leading to a 3/2-bit/s/Hz transmission per user. 
Figure \ref{BER_figure_2} extends the comparison between SM-encoded MAT and Alternative MAT and TDMA at higher rates. Note that the displayed SNR on the x-axis is $\rho$. Hence for a SNR of $\rho$, TDMA is allocated a power $4/3\rho$ in order to keep the total average transmit power the same for MAT and TDMA.
The behaviour follows the observations made from the analytical results on diversity and coding gains of TDMA with O-STBC and MAT/Alt MAT with SM/full rank codes (Section \ref{iid_error_rate_section}). It also confirms that TDMA has a larger diversity gain than SM-encoded MAT and Alternative MAT but is impacted by the use of larger constellation sizes. Namely, we make the following observations: 1) At low rate (3 bit/s/Hz per user and below), MAT outperforms TDMA at high SNR if it is concatenated with a full-rate full-diversity space-time code (see Figure \ref{BER_figure}), 2) SM-encoded MAT does not show any significant benefit over TDMA in the simulated rate and SNR range (Figures \ref{BER_figure} and \ref{BER_figure_2}), 3) Alternative MAT does not show any benefit over TDMA in the simulated rate and SNR range (Figures \ref{BER_figure} and \ref{BER_figure_2}). At even higher rate (above 3 bit/s/Hz per user), SM-encoded MAT (i.e.\ without requiring any additional encoding in the spatial domain) is expected to start showing some meaningful performance benefits over TDMA at low SNR.

\par
\begin{figure}
\centerline{\includegraphics[width=\columnwidth]{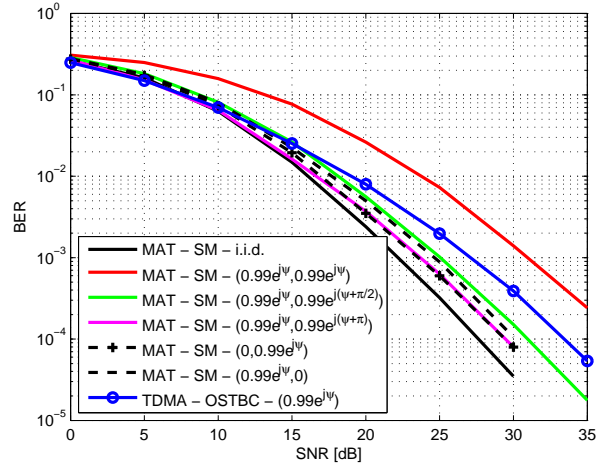}}
  \caption{Average Bit Error Rate (BER) of SM-encoded MAT (with R=4/3 bit/s/Hz per user) over spatially correlated channels $\big(t_1,t_2\big)$. Phase $\psi$ is random and uniformly distributed within $\left[0,2\pi\right]$. Comparison with TDMA - OSTBC (R=3/2 bit/s/Hz) is also provided.}
  \label{BER_MAT_correle_random_phases_figure}
\end{figure}

 In Figure \ref{BER_MAT_correle_random_phases_figure}, the BER performance of user 1 with QPSK-based SM-encoded MAT (i.e.\ R=4/3 bit/s/Hz per user) in spatially correlated channels is displayed for various pairs of the transmit correlation coefficients  $(t_1,t_2)$. The performance of TDMA with OSTBC at a rate per user of R=3/2 bit/s/Hz is also provided. In the presence of high transmit correlation, the phase of the correlation coefficient is indicative of the location of the user w.r.t.\ the transmit array as $\varphi\approx 2\pi d/\lambda \cos \theta$ with $d$ the inter-element spacing and $\theta$ the angle of departure (taken w.r.t.\ the array axis). Increasing the transmit correlation coefficients phase shift (i.e.\ $\phi=\varphi_2-\varphi_1$) makes the users' channels more statistically orthogonal. In the evaluations, $\psi$ is random and uniformly distributed within $\left[0,2\pi\right]$ such that the scenario $(t_1,t_2)=(0.99e^{j\psi},0.99e^{j(\psi+\phi)})$ refers to the case where $\left|t_1\right|=\left|t_2\right|=0.99$, $\varphi_1=\psi$ is randomly distributed and $\varphi_2=\varphi_1+\phi$. We make the following observations, inline with the analytical derivations: 
\begin{enumerate}
\item The performance improves as users get statistically orthogonal to each other. By decreasing order of BER, we have $\phi$ equal to 0,$\pi/2$,$\pi$.  
\item I.i.d.\ channels and spatially correlated channels with statistically orthogonal users $\left(0.99e^{j\psi},0.99e^{j(\psi+\pi)}\right)$ lead to similar performance at low to medium SNR (curves are superposed up to 15dB).
\item Pairing two statistically orthogonal users with similar magnitudes of the transmit correlation coefficients outperforms pairing two users with asymmetric spatial correlation (one with large spatial correlation and the other one with low spatial correlation). This can be seen by comparing $\left(0.99e^{j\psi},0.99e^{j(\psi+\pi)}\right)$ with $\left(0,0.99e^{j\psi}\right)$ and $\left(0.99e^{j\psi},0\right)$. Hence, interestingly, high transmit spatial correlation can lead to a better performance than low transmit spatial correlation. This contrasts with the conventional impact of transmit correlation on error rate performance in point-to-point channels \cite{Clerckx:2007b}.
\item MAT is shown to be less sensitive to spatial correlation than TDMA - OSTBC. Indeed, SM-encoded MAT outperforms TDMA in spatially correlated channels with proper user pairing (e.g.\ $\phi=\pi/2,\pi$). This contrasts with the behavior on i.i.d. channels where SM-encoded MAT is outperformed by TDMA at the same rates (see Figure \ref{BER_figure}).
\end{enumerate}


\par
\begin{figure}
\centerline{\includegraphics[width=\columnwidth]{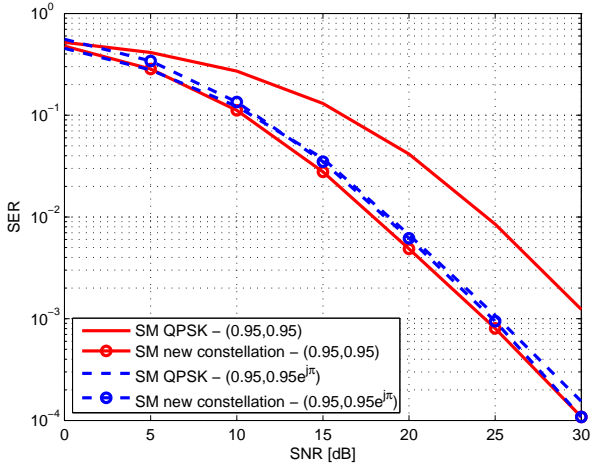}}
  \caption{Average Symbol Error Rate (SER) of SM-encoded MAT over spatially correlated channels $\big(t_1,t_2\big)$ with QPSK and optimized constellations.}
  \label{SER_MAT_correle_new}
\end{figure}

 We now investigate the performance gain of non-linear signal constellations $\mathbf{S}$ and $\left\{\mathbf{Q}_m\right\}$ designed to replace conventional QPSK in a 4/3-bit/s/Hz transmission. They
have been optimized for several pairs of ($t_1$,$t_2$), whereas $\rho/4$ is chosen equal to 20dB. A large number of initial
conditions have been tested, and less than 1000 iterations
were necessary to converge to an optimum. Several values of $\alpha$ in \eqref{constrained_gradient}
have been considered, depending on the
speed of convergence of the algorithm. The constellations that have been used in Figure \ref{SER_MAT_correle_new} are those that provide the minimum $\bar{P}$
among the encountered local optimal constellations. In the configuration $\left(0.95,0.95e^{j\pi}\right)$, optimized constellations get closer to QPSK constellations as $\mathbf{R}_{t,e}$ gets closer to a scaled identity matrix. Contrary to the QPSK constellations in Figure \ref{BER_MAT_correle_random_phases_figure}, the adaptive non-linear constellations make the performance almost insensitive to the phase shift $\varphi_2-\varphi_1$. Indeed, curves are almost superposed to each other at high SNR (recall the optimization is made for $\rho/4$ equal to 20dB). Hence the transmitter does not have to worry about user orthogonality and could schedule any pair of users irrespectively of their orthogonality.

\section{Conclusions}\label{conclusions}

The error rate performance of space-time encoded MISO BC with delayed CSIT has been analyzed. The paper provides new insights into the diversity-multiplexing tradeoff achievable with various transmissions strategies and derives novel space-time code designs, signal constellations and user pairing strategies for multi-user communications in the presence of outdated CSIT. The effect of transmit spatial correlation on the performance is also highlighted and the performance behaviour is contrasted with conventional MU-MISO with perfect CSIT and point-to-point MIMO channels.
The paper contrasts with the common DoF analysis and provides new insights into the actual error rate performance benefits over TDMA. The main takeaway message is that in i.i.d.\ Rayleigh fading channels, MAT is not always a better alternative than TDMA. Benefits are obtained over TDMA (based on O-STBC) only at very high rate or once concatenated with a full-rate full-diversity space-time code. Alternative MAT is even less competitive than MAT and is outperformed by TDMA over a wide SNR and rate range. In spatially correlated channels, MAT with a suitable user pairing strategy is shown to be less sensitive to transmit spatial correlation than TDMA. 

\appendices

\section{Average PEP of Space-Time Encoded MAT}\label{Appendix_MAT}
The average PEP is obtained by taking the expectation of the conditional PEP \eqref{MAT_PEP_cond} over the channel distribution. The expectation can be computed in two steps: first by taking the expectation over the distribution of $\mathbf{H}'$ and then over the distribution of $X$. The first step writes as
\begin{align}
P\left(\mathbf{C}\rightarrow \mathbf{E}\left|\right.X\right)&=\mathcal{E}_{\mathbf{H}'}\left\{P\left(\mathbf{C}\rightarrow \mathbf{E}\left|\right.\tilde{\mathbf{H}}\right)\right\}\nonumber\\
&=\left(\det\left(\mathbf{I}_4+\frac{\rho}{4}\mathbf{R}\left[\mathbf{I}_2\otimes \tilde{\mathbf{E}}\right]\right)\right)^{-1}\label{MAT_PEP_first_step}
\end{align}
where 
\begin{equation}
\mathbf{R}=\mathcal{E}\left\{\textnormal{vec}\left(\tilde{\mathbf{H}}^H\right)\textnormal{vec}\left(\tilde{\mathbf{H}}^H\right)^H\right\}=\textnormal{diag}\left\{\mathbf{R}_{t,1},\left|X\right|^2\mathbf{R}_{t,2}\right\}
\end{equation}
and $\tilde{\mathbf{E}}=\left(\mathbf{C}-\mathbf{E}\right)\left(\mathbf{C}-\mathbf{E}\right)^H$ is the error matrix. The component block matrices $\mathbf{R}_{t,1}$ and $\mathbf{R}_{t,2}$ are defined as the transmit correlation matrices for receiver 1 and 2 as in \eqref{correlation_matrices}. 

Expression \eqref{MAT_PEP_first_step} is obtained from the classical derivation of average PEP over MIMO spatially correlated Rayleigh fading channels \cite{Clerckx:2013,Clerckx:2007b}.
We can further expand as
\begin{align}
&P\left(\mathbf{C}\rightarrow \mathbf{E}\left|\right.X\right)\nonumber\\
&\hspace{0.4cm}=\left(\det\left(\mathbf{I}_2+\frac{\rho}{4}\mathbf{R}_{t,1}\tilde{\mathbf{E}}\right)\right)^{-1}\left(\det\left(\mathbf{I}_2+\frac{\rho}{4}\left|X\right|^2\mathbf{R}_{t,2}\tilde{\mathbf{E}}\right)\right)^{-1}\nonumber\\
&\hspace{0.4cm}= b_{1,1}^{-1}b_{2,1}^{-1}\left(1+a_{1,2}\left|X\right|^2\right)^{-1}\left(1+a_{2,2}\left|X\right|^2\right)^{-1}
\end{align}
where $b_{k,i}=1+a_{k,i}$ ($k,i=1,2$) with $a_{k,i}=\frac{\rho}{4}\lambda_{k,i}$ and $\lambda_{k,i}$ the $k^{th}$ eigenvalue of the matrix $\mathbf{R}_{t,i}\tilde{\mathbf{E}}$. We can moreover write $\left|X\right|^2=\frac{Y}{1+Y}$ where $Y=\left|h_{3,1}\right|^2$ is exponentially distributed. In the particular case of i.i.d.\ Rayleigh fading, $b_{k,1}=b_{k,2}=b_k=1+a_k$ ($k=1,2$) with $a_k=\frac{\rho}{4}\lambda_k$ and $\lambda_k$ the $k^{th}$ eigenvalue of the error matrix $\tilde{\mathbf{E}}$.

As a second step, we take the expectation over the distribution of $X$ (or equivalently $Y$). We write
\begin{multline}
P\left(\mathbf{C}\rightarrow \mathbf{E}\right)=\mathcal{E}_{X}\left\{P\left(\mathbf{C}\rightarrow \mathbf{E}\left|\right.X\right)\right\}=b_{1,1}^{-1}b_{2,1}^{-1}\\
\int_0^\infty\left(1+a_{1,2}\frac{y}{1+y}\right)^{-1}\left(1+a_{2,2}\frac{y}{1+y}\right)^{-1}e^{-y}dy.\label{MAT_PEP_second_step}
\end{multline}
Making use of partial fraction expansion and Table of Integrals \cite{Gradshteyn:2007}, the integral in \eqref{MAT_PEP_second_step} can be solved and we obtain \eqref{MAT_PEP} in i.i.d.\ Rayleigh fading channels and \eqref{MAT_PEP_correlated} in spatially correlated Rayleigh fading channels. $\textnormal{Ei}\left(x\right)=\int_{-\infty}^x\frac{e^t}{t}dt$ in \eqref{MAT_PEP} and \eqref{MAT_PEP_correlated} is the exponential integral whose Taylor series for real argument write as $\textnormal{Ei}\left(x\right)=\gamma+\log \left|x\right|+\sum_{i=1}^{\infty}\frac{x^i}{i \hspace{0.1cm}i!}$ where $\gamma$ is the Euler-Mascheroni constant.

\par At high SNR, assuming a full-rank code, $a_{k,i}>>0$ and $b_{k,i}>>0$ $\forall k,i$ and for small argument $x$, $\textnormal{Ei}\left(x\right)\approx \gamma+\log \left|x\right|$. Applying those simplifications at high SNR to \eqref{MAT_PEP} and \eqref{MAT_PEP_correlated} and keeping only the dominant terms lead to the approximations \eqref{MAT_PEP_highSNR} and \eqref{MAT_PEP_highSNR_correlated}. The PEP expressions for SM and O-STBC in i.i.d.\ channels \eqref{MAT_PEP_SM} and \eqref{MAT_PEP_OSTBC} and in spatially correlated channels \eqref{MAT_PEP_SM_correlated} are obtained from \eqref{MAT_PEP} and \eqref{MAT_PEP_correlated} by directly plugging the corresponding parameters. 
In i.i.d.\ channels, the PEP for O-STBC (for which $b_1=b_2=b$) involves the computation of $\lim_{b_1\rightarrow b_2} \frac{f\left(b_1\right)-f\left(b_2\right)}{b_1-b_2}$ with $f(x)=\frac{\left(x-1\right)^2}{x^2}e^{1/x}\textnormal{Ei}\left(\frac{-1}{x}\right)$. This can be done by noting that $\frac{\partial \textnormal{Ei}\left(x\right)}{\partial x}=\frac{e^x}{x}$.

\section{Proof of Theorem \ref{Theorem_MAT}}\label{proof_DMT_MAT}
Following \cite{Chen:2012} and the system model \eqref{system_model_MAT}, we can write the outage probability
\begin{equation}
P_{\textnormal{out}}(r)\stackrel{.}{=}P\left(\log_2\det\left(\mathbf{I}_2+\frac{\rho}{2}\mathbf{H}\mathbf{H}^H\right)<3R\right)
\end{equation}
where $R=r \log_2 \rho$.
Making use of the QR decomposition $\mathbf{H}=\mathbf{Q}\mathbf{R}$ and $\mathbf{H}'=\mathbf{Q}'\mathbf{R}'$ with $(i,j)$ entry of $\mathbf{R}$ (resp.\ $\mathbf{R}'$) denoted as $r_{ij}$ ($i,j=1,2$) and $r_{21}=0$ (resp.\ $r_{ij}'$ with $r_{21}'=0$), we get
\begin{align}\label{Pout}
P_{\textnormal{out}}(r)&\stackrel{.}{=}P\left(1+\frac{\rho}{2}\left\|\mathbf{H}\right\|_F^2+\left(\frac{\rho}{2}\right)^2\left|\det\left(\mathbf{H}\right)\right|^2<\rho^{3r}\right)
\end{align}
where $\left\|\mathbf{H}\right\|_F^2=r_{11}^2+\left|r_{12}\right|^2+r_{22}^2$ and $\left|\det\left(\mathbf{H}\right)\right|^2=r_{11}^2r_{22}^2$. Since $\det\left(\mathbf{H}\right)=h_{3,1}\det\left(\mathbf{H}'\right)$, we also have $\left|\det\left(\mathbf{H}\right)\right|^2=\left|h_{3,1}\right|^2 r'^{2}_{11} r'^{2}_{22}$.

\par For $\frac{1}{3}\leq r\leq \frac{2}{3}$, $1+\frac{\rho}{2}\left\|\mathbf{H}\right\|_F^2<< \rho^{3r}$ such that
\begin{equation} P_{\textnormal{out}}(r)\stackrel{.}{=}P\left(\left(\frac{\rho}{2}\right)^2\left|h_{3,1}\right|^2 r'^{2}_{11} r'^{2}_{22}<\rho^{3r}\right).
\end{equation} 
Similarly to \cite{Yao:2003}, we can now use the fact that the dominant error event is given by $r'^{2}_{11}<1$ and $\left|h_{3,1}\right|^2r'^{2}_{22}<\rho^{3r-2}$ because $r'^{2}_{11}$ is likely to be larger than $\left|h_{3,1}\right|^2r'^{2}_{22}$. Indeed $r'^{2}_{11}\sim\chi_4^2$ while $\left|h_{3,1}\right|^2r'^{2}_{22}$ writes as the product of two independent $\chi_2^2$. 
\par Assuming $X,Y\sim\chi_2^2$ and independent, we can compute the CDF of $U=XY$ as
\begin{equation}
P\left[U\leq \epsilon\right]=1-\int_0^{\infty}e^{-\left(\frac{\epsilon}{v}+v\right)}dv=1-2\sqrt{\epsilon}K_1(2\sqrt{\epsilon})
\end{equation}
where $K_1(x)$ is the modified Bessel function of the second kind. The second equality comes from \cite{Gradshteyn:2007}.
Assuming small $x$, from the series expansions in \cite{Abramovitz:1972}, $x K_1(x)\approx1+x^2 \log\left(x/2\right)$. Hence, for small $\epsilon$, $P\left[U\leq \epsilon\right]\approx-2\epsilon\log(\epsilon)$.
\par The outage probability then writes at large SNR as
\begin{align} P_{\textnormal{out}}(r)&\stackrel{.}{=}P\left(r'^{2}_{11}<1 \:\textnormal{and}\:\left|h_{3,1}\right|^2 r'^{2}_{22}<\rho^{3r-2}\right)\nonumber\\
&\stackrel{.}{=}-2\rho^{3r-2}\log\left(\rho^{3r-2}\right)\stackrel{.}{=}\rho^{-(2-3r)}.
\end{align}
Therefore, $d^{\star}(r)=2-3r$
for $\frac{1}{3}\leq r\leq \frac{2}{3}$. Hence the piecewise-linear function joining $(\frac{1}{3},1)$ and $(\frac{2}{3},0)$. 

\par For $0\leq r\leq \frac{1}{3}$, 
\begin{multline}\label{Pout_medium}
P_{\textnormal{out}}(r)\stackrel{.}{=}P\bigg(\frac{\rho}{2}\left(\left|h_{3,1}\right|^2\left[\left|g_{1,1}\right|^2+\left|g_{1,2}\right|^2\right]+\left|h_{1,1}\right|^2+\left|h_{1,2}\right|^2\right)\bigg.\\\bigg.+\left(\frac{\rho}{2}\right)^2\left|h_{3,1}\right|^2 r'^{2}_{11} r'^{2}_{22}<\rho^{3r}\bigg)
\end{multline}
and we can conclude from the first order term that the dominant outage events are such that $\left|h_{1,1}\right|^2<\rho^{3r-1}$, $\left|h_{1,2}\right|^2<\rho^{3r-1}$, $\left|h_{3,1}\right|^2<\rho^{3r-1}$, $\left|g_{1,1}\right|^2<1$ and $\left|g_{1,2}\right|^2<1$. The last three inequalities come from the fact that $\left|h_{3,1}\right|^2\big[\left|g_{1,1}\right|^2+\left|g_{1,2}\right|^2\big]$ is likely to be smaller than $\rho^{3r-1}$ whenever $\left|h_{3,1}\right|^2<\rho^{3r-1}$ because $\left|h_{3,1}\right|^2\sim\chi_2^2$ while $\left|g_{1,1}\right|^2+\left|g_{1,2}\right|^2\sim\chi_4^2$. Those inequalities imply that
$\left|\det\left(\mathbf{H}'\right)\right|^2=r'^{2}_{11} r'^{2}_{22}\stackrel{.}{<}\rho^{3r-1}$.
Therefore, in order to guarantee $\left|h_{3,1}\right|^2 r'^{2}_{11} r'^{2}_{22}<\rho^{3r-2}$ in the second order term of \eqref{Pout_medium}, $\left|h_{3,1}\right|^2$ should be even smaller, less than $\rho^{-1}$. The outage probability writes as
\begin{align}\label{P_out_low_r}
P_{\textnormal{out}}(r)&\stackrel{.}{=}P\left(\left|h_{3,1}\right|^2<\rho^{-1} \:\textnormal{and}\: \left|h_{1,1}\right|^2<\rho^{3r-1}\:\textnormal{and}\:\right.\nonumber\\
&\hspace{0.5cm}\left. \left|h_{1,2}\right|^2<\rho^{3r-1}\:\textnormal{and}\: \left|g_{1,1}\right|^2<1\:\textnormal{and}\: \left|g_{1,2}\right|^2<1\right)\nonumber\\
&\stackrel{.}{=} \rho^{-1}\rho^{3r-1}\rho^{3r-1}\stackrel{.}{=}\rho^{-(3-6r)}.
\end{align}
Hence, $d^{\star}(r)=3-6r$ for $0\leq r\leq \frac{1}{3}$, i.e.\ the piecewise-linear function joining $(0,3)$ and $(\frac{1}{3},1)$. From \eqref{P_out_low_r}, the dominant outage event occurs whenever the three channel coefficients $h_{3,1}$, $h_{1,1}$ and $h_{1,2}$ are in deep fade. $g_{1,1}$ and $g_{1,2}$ do not influence the DMT as both are multiplied by $h_{3,1}$ in \eqref{H_system_model_MAT}. 

\section{Proof of Theorem \ref{Theorem_MAT_SM} and \ref{Theorem_AltMAT_SM}}\label{proof_DMT_SM_MAT_altMAT}
From \eqref{SM_highSNR_PEP_MAT} and \eqref{SM_highSNR_PEP_AltMAT}, the average PEP with SM at high SNR writes as $P\left(\mathbf{C}\rightarrow \mathbf{E}\right)\leq \frac{\left(\log\left(a\right)\right)^m}{a^2}$ with $m=1$ for MAT and $m=2$ for Alternative MAT. Assuming without loss of generality $\mathbf{E}=\mathbf{0}$, we can bound the overall error probability using the union bound
\begin{align}
P_e&\leq \sum_{\mathbf{C}\neq \mathbf{0}}P\left(\mathbf{C}\rightarrow \mathbf{0}\right)\nonumber\\
&\leq\sum_{\mathbf{C}\neq \mathbf{0}}\left(\frac{\rho}{4}\right)^{-2}\left(\sum_{q=1}^{2} \left|c_q\right|^2\right)^{-2}\left(\log\left(\frac{\rho}{4}\sum_{q=1}^{2} \left|c_q\right|^2\right)\right)^{m}.
\end{align}
Assuming that the same data rate $R_q=3R/2$ is assigned to both streams (with $R=r\log\rho$), symbols $c_q$ are chosen from a QAM constellation carrying $2^{R_q/2}$ symbols per dimension. For a unit average energy, the minimum distance $d_{min}$ between two points of such a constellation is of the order $1/{2^{R_q/2}}$.
We write $c_q=\left(i_{2q-1}+ji_{2q}\right)d_{min}$ where $i_x\in\mathbbm{Z}$ ($x=1,...,4$). The minimum squared distance of the constellation is of the order $d_{min}^2=2^{-3R/2}=\rho^{-3r/2}$ and we can further bound $P_e$ as 
\begin{align}
P_e&\leq\sum_{\mathbf{C}\neq \mathbf{0}}4^2\rho^{-(2-3r)}\frac{\left(\log\left(\frac{\rho^{1-3r/2}}{4}\sum_{x=1}^{4} \left|i_x\right|^2\right)\right)^{m}}{\left(\sum_{x=1}^{4}\left|i_x\right|^2\right)^{2}} \nonumber\\
&=4^2\rho^{-(2-3r)}\sum_{\left(i_1,\ldots,i_4\right)\neq \mathbf{0}}\frac{A^m+B^m+2(m-1)AB}{C^{2}}
\end{align}
where $A=\log\left(\frac{\rho^{1-3r/2}}{4}\right)$, $B=\log C$ and $C=\left|i_1\right|^2+\ldots+\left|i_4\right|^2$.
Following \cite{Tavildar:2006}, 
\begin{equation}
\sum_{\left(i_1,\ldots,i_4\right)\neq \mathbf{0}}\frac{1}{C^{2}}\stackrel{.}{=}1.
\end{equation}
Similarly, from Appendix \ref{Appendix_proof_SM_series}, we can compute
\begin{equation}\label{proof_sum}
\sum_{\left(i_1,\ldots,i_4\right)\neq \mathbf{0}}\frac{B^{m}}{C^{2}}\stackrel{.}{=}1.
\end{equation}
This leads to 
\begin{align}
P_e&\stackrel{.}{\leq}\rho^{-(2-3r)}\left(\left(\log\left(\rho^{1-3r/2}\right)\right)^m+(m-1)\log\left(\rho^{1-3r/2}\right)\right)\nonumber\\
&\stackrel{.}{=}\rho^{-(2-3r)}.
\end{align}
Hence, the DMT achieved by SM-encoded MAT and Alternative MAT is given by $d(r)=2-3r$ for $r\in\left[0,\frac{2}{3}\right]$.

\section{Proof of \eqref{proof_sum}}\label{Appendix_proof_SM_series}
Similarly to \cite{Tavildar:2006}, the summation is split over vectors with nonzero components and is successively upper bounded using several inequalities. In the sequel, $S$ denotes a subset of $\left\{1,2,3,4\right\}$. We write the following inequalities
\begin{align}
&\sum_{\left(i_1,\ldots,i_4\right)\neq \mathbf{0}}\frac{B^{m}}{C^{2}}\nonumber\\
&=\sum_S \sum_{i_x\neq 0:x\in S}\frac{\left(\log\left(\left|i_1\right|^2+\ldots+\left|i_4\right|^2\right)\right)^{m}}{\left(\left|i_1\right|^2+\ldots+\left|i_4\right|^2\right)^{2}}\nonumber\\
&= \sum_S \sum_{i_x\neq 0:x\in S}\frac{\left(\log\left(\sum_{x\in S}\left|i_x\right|^2\right)\right)^{m}}{\left(\sum_{x\in S}\left|i_x\right|^2\right)^{2}}\nonumber\\
&\stackrel{(\textnormal{a})}{\leq} \sum_S \sum_{i_x\neq 0:x\in S}\frac{\left(\sum_{x\in S}\left|i_x\right|^2\log\left(\left|i_x\right|^2\right)\right)^{m}}{\left(\sum_{x\in S}\left|i_x\right|^2\right)^{2+m}}\nonumber\\
&\stackrel{(\textnormal{b})}{\leq} \sum_S \sum_{i_x\neq 0:x\in S}\frac{\left(\sum_{x\in S}\left|i_x\right|^2\log\left(\left|i_x\right|^2\right)\right)^{m}}{\prod_{x\in S}\left|i_x\right|^{2(2+m)/\left|S\right|}}\nonumber\\
&\leq \sum_S \sum_{i_x\neq 0:x\in S} \left[\sum_{x\in S}\frac{\left(\log\left(\left|i_x\right|^2\right)\right)^m}{\left|i_x\right|^{4/\left|S\right|}}\right.\nonumber\\
&\hspace{0.5cm}\left.+(m-1)\sum_{\mycom{x_1,x_2\in S}{x_1\neq x_2}}\frac{\log\left(\left|i_{x_1}\right|^2\right)\log\left(\left|i_{x_2}\right|^2\right)}{\left|i_{x_1}\right|^{2(1+m)/\left|S\right|}\left|i_{x_2}\right|^{2(1+m)/\left|S\right|}}\right]\nonumber\\
&\leq \sum_S \sum_{i_x\neq 0:x\in S} \left[\sum_{x\in S}\frac{\left(\log\left(\left|i_x\right|^2\right)\right)^m}{\left|i_x\right|^{4/\left|S\right|}}\right.\nonumber\\
&\hspace{0.5cm}\left.+(m-1)\sum_{\mycom{x_1,x_2\in S}{x_1\neq x_2}}\frac{1}{\left|i_{x_1}\right|^{2m/\left|S\right|}\left|i_{x_2}\right|^{2m/\left|S\right|}}\right]\nonumber\\
&\leq \sum_S \sum_{i_x\neq 0:x\in S} \left[\sum_{x\in S}\frac{\left(\log\left(\left|i_x\right|^2\right)\right)^m}{\left|i_x\right|^{4/\left|S\right|}}\right.\nonumber\\
&\hspace{0.5cm}\left.+(m-1)\left(\left|S\right|-1\right)\sum_{x\in S}\frac{1}{\left|i_{x}\right|^{2m/\left|S\right|}}\right]\nonumber\\
&= \sum_S \left|S\right|\left[\sum_{i_1\neq 0} \frac{\left(\log\left(\left|i_1\right|^2\right)\right)^m}{\left|i_1\right|^{4/\left|S\right|}}\right.\nonumber\\
&\hspace{0.5cm}\left.+(m-1)\left(\left|S\right|-1\right)\sum_{i_1\neq 0}\frac{1}{\left|i_{1}\right|^{2m/\left|S\right|}}\right]
\end{align}
where (a) results from the application of the log sum inequality \cite{Cover:2006} and (b) from the inequality of arithmetic and geometric means. Since $\left|S\right|\leq 4$, and recalling that $\left|i_1\right|$ grows with the SNR,
\begin{equation}
\sum_{i_1\neq 0} \frac{\left(\log\left(\left|i_1\right|^2\right)\right)^m}{\left|i_1\right|^{4/\left|S\right|}}\stackrel{.}{=}1 \hspace{0.2cm}\textnormal{and}\hspace{0.2cm}\sum_{i_1\neq 0}\frac{1}{\left|i_{1}\right|^{4/\left|S\right|}}\stackrel{.}{=}1,\nonumber
\end{equation}
and we finally obtain $\sum_{\left(i_1,\ldots,i_4\right)\neq \mathbf{0}}\frac{B^{m}}{C^{2}}\stackrel{.}{=}1$.

\section{Analytical Expression of the Gradients}\label{Appendix_gradient}
The gradients of $\bar{P}$ with respect to $S_k$ and $Q_{xk}$ are respectively expressed as displayed in \eqref{gSk} and \eqref{gQxk},
\begin{table*}
\begin{multline}\label{gSk}
g_{S_k}=\nabla_{S_k} \bar{P}=\frac{-2}{M_0 M_1} \sum_{u\neq k}^{M_0}\sum_{n=1}^{M_1}\sum_{v=1}^{M_1}\frac{\frac{\rho}{4} \left(2\left(S_k-S_u\right)+2t_1^{*}\left(Q_{kn}-Q_{uv}\right)\right)s\left(S_k,Q_{kn},S_u,Q_{uv}\right)}{\left(1+\frac{\rho}{4}A\right)^2\left(1+\frac{\rho}{4}B\right)}\\+\frac{\frac{\rho}{4} \left(2\left(S_k-S_u\right)+2t_2^{*}\left(Q_{kn}-Q_{uv}\right)\right)s\left(S_k,Q_{kn},S_u,Q_{uv}\right)}{\left(1+\frac{\rho}{4}A\right)\left(1+\frac{\rho}{4}B\right)^2}
\end{multline}
\begin{multline}\label{gQxk}
g_{Q_{xk}}=\nabla_{Q_{xk}} \bar{P}=\frac{-2}{M_0 M_1} \sum_{u=1}^{M_0}\sum_{v=1}^{M_1}
\frac{\frac{\rho}{4} \left(2\left(Q_{xk}-Q_{uv}\right)+2t_1\left(S_x-S_u\right)\right)s\left(S_x,Q_{xk},S_u,Q_{uv}\right)}{\left(1+\frac{\rho}{4}C\right)^2\left(1+\frac{\rho}{4}D\right)}\\
+\frac{\frac{\rho}{4} \left(2\left(Q_{xk}-Q_{uv}\right)+2t_2\left(S_x-S_u\right)\right)s\left(S_x,Q_{xk},S_u,Q_{uv}\right)}{\left(1+\frac{\rho}{4}C\right)\left(1+\frac{\rho}{4}D\right)^2}
\end{multline}
\hrulefill
\end{table*}
where
\begin{align}
\begin{split}
A&=\left|S_k-S_u\right|^2+\left|Q_{kn}-Q_{uv}\right|^2\\
&\hspace{1cm}+2\Re\left\{t_1\left(S_k-S_u\right)\left(Q_{kn}-Q_{uv}\right)^*\right\},\\
B&=\left|S_k-S_u\right|^2+\left|Q_{kn}-Q_{uv}\right|^2\\
&\hspace{1cm}+2\Re\left\{t_2\left(S_k-S_u\right)\left(Q_{kn}-Q_{uv}\right)^*\right\},\\
C&=\left|S_x-S_u\right|^2+\left|Q_{xk}-Q_{uv}\right|^2\\
&\hspace{1cm}+2\Re\left\{t_1\left(S_x-S_u\right)\left(Q_{xk}-Q_{uv}\right)^*\right\},\\
D&=\left|S_x-S_u\right|^2+\left|Q_{xk}-Q_{uv}\right|^2\\
&\hspace{1cm}+2\Re\left\{t_2\left(S_x-S_u\right)\left(Q_{xk}-Q_{uv}\right)^*\right\}.
\end{split}
\end{align}

\section{Average PEP of Space-Time Encoded Alternative MAT}\label{Appendix_Alt_MAT}
Taking the expectation of the conditional pairwise error probability (PEP) $P\big(\mathbf{C}\rightarrow \mathbf{E}\left|\right.\tilde{\mathbf{H}}\big)$ \eqref{MAT_PEP_cond} over the distribution of $\mathbf{H}'$ assuming $Z$ and $X$ fixed leads to 
\begin{align}
P\left(\mathbf{C}\rightarrow \mathbf{E}\left|\right.X,Z\right)&=\mathcal{E}_{\mathbf{H}'}\left\{P\left(\mathbf{C}\rightarrow \mathbf{E}\left|\right.\tilde{\mathbf{H}}\right)\right\}\nonumber\\
&=\left(\det\left(\mathbf{I}_4+\frac{\rho}{4}\mathbf{R}\left[\mathbf{I}_2\otimes \tilde{\mathbf{E}}\right]\right)\right)^{-1}\label{Alt_MAT_PEP_first_step}
\end{align}
where 
\begin{align}
\mathbf{R}&=\mathcal{E}\left\{\textnormal{vec}\left(\tilde{\mathbf{H}}^H\right)\textnormal{vec}\left(\tilde{\mathbf{H}}^H\right)^H\right\}\nonumber\\
&=\textnormal{diag}\left\{\left|Z\right|^2\mathbf{R}_{t,2},\left|X\right|^2\mathbf{R}_{t,1}\right\}
\end{align}
with the transmit correlation matrices defined as in \eqref{correlation_matrices}.

Alternatively, we can write the conditional PEP on $X$ and $Z$ as in \eqref{Alt_MAT_appendix_alternative}
\begin{table*}
\begin{align}
P\left(\mathbf{C}\rightarrow \mathbf{E}\left|\right.X,Z\right)
&=\left(\det\left(\mathbf{I}_2+\frac{\rho}{4}\left|Z\right|^2\mathbf{R}_{t,2}\tilde{\mathbf{E}}\right)\right)^{-1}\left(\det\left(\mathbf{I}_2+\frac{\rho}{4}\left|X\right|^2\mathbf{R}_{t,1}\tilde{\mathbf{E}}\right)\right)^{-1}\nonumber\\
&=\prod_{k=1,2}\left(1+a_{k,2}\left|Z\right|^2\right)^{-1}\left(1+a_{k,1}\left|X\right|^2\right)^{-1},\label{Alt_MAT_appendix_alternative}
\end{align}
\hrulefill
\end{table*}
where $a_{k,i}$ and $b_{k,i}$ are defined as in Appendix \ref{Appendix_MAT}. Integrating over the distribution of X and Z, we obtain
\begin{multline}
P\left(\mathbf{C}\rightarrow \mathbf{E}\right)=\int_0^\infty\prod_{k=1}^2\left(1+a_{k,2} y\right)^{-1}e^{-y}dy\\\int_0^\infty\prod_{k=1}^2\left(1+a_{k,1}\frac{y}{1+y}\right)^{-1}e^{-y}dy.\label{Alt_MAT_PEP_second_step}
\end{multline}
Making use of partial fraction expansion and Table of Integrals \cite{Gradshteyn:2007}, integrals can be solved and we obtain \eqref{Alt_MAT_PEP} in i.i.d.\ channels and \eqref{Alt_MAT_PEP_correlated} in spatially correlated channels.


\ifCLASSOPTIONcaptionsoff
  \newpage
\fi

\end{document}